\documentclass[11pt,a4paper]{article}

\usepackage{amsmath,amssymb}
\usepackage{epsfig,graphicx}
\usepackage{subfigure}
\usepackage{graphicx}
\usepackage{rotating}
\usepackage{cancel}
\usepackage{bm}
\usepackage{color}
\usepackage{comment}
\usepackage{cite}
\usepackage{psfrag}
\usepackage{slashed}
\usepackage{soul}
\usepackage{tikz}
\usepackage{hyperref}

\newcommand{\Nspin}{N_{s}}
\newcommand{\Naxion}{N_{a}}
\newcommand{\Nup}{N_{\uparrow}}
\newcommand{\Ndown}{N_{\downarrow}}
\newcommand{\oc}{\omega_{c}}

\renewcommand\[{\left[}

\newcommand{\exclude}[1]{}

\def\bra{\langle}
\def\ket{\rangle}
\def\beq{\begin{equation}}
\def\eeq{\end{equation}}

\begin{document}
\title{\Large{\textbf{A Quantum Perspective on Oscillation Frequencies}} \\ \Large{ \textbf{in Axion Dark Matter Experiments}}}

\author{Joerg Jaeckel, Valentina Montoya and Cedric Quint\\[2ex]
\small{\em Institut f\"ur theoretische Physik, Universit\"at Heidelberg,} \\
\small{\em Philosophenweg 16, 69120 Heidelberg, Germany}\\[0.5ex]}

\date{}
\maketitle

\begin{abstract}
\noindent
In this note we look at the time evolution of signals in axion dark matter experiments from a quantum perspective. Our aim is not to contribute new results to the general discussion of the quantum/classical connection (which we do not) but rather to slightly illuminate the specific case of axion experiments.
From the classical perspective one expects a signal oscillating with a frequency equal to the axion mass whose amplitude is slowly rising due to the tiny interaction of the axions with ordinary matter. In the quantum picture the latter, slow time-scale arises from the small splitting in the energy levels induced by the interaction between the axions and the experiment, and it is always present in suitable, sensitive experiments. Signals that oscillate with a frequency equal to the axion mass, however, arise from processes changing the axion number. Yet, depending on the chosen observable, such oscillations may be absent for certain special initial quantum states of the axions. However, we show by example that, using an appropriate experimental procedure, these special states can be modified by the experiment in such a way that a signal oscillating with the axion mass re-appears. 
In addition, we discuss the measurement of suitable correlators that feature an oscillation with the axion mass.
We also comment on the connection to the classical treatment.
The explicit experiment we look at is an oscillating EDM experiment such as CASPEr but we expect our results to be easily adaptable to other types of axion dark matter experiments.

\end{abstract}

\newpage

\tableofcontents
\newpage


\section{Introduction}\label{sec:intro}
To describe the effects of dark matter axions in experiments it is common (and indeed quite reasonable) to treat them as a classical field. Justification for this is provided by the typically enormous occupation numbers $N\sim 10^{41}({\rm neV}/m_{a})^4$~\cite{Sikivie:1983ip} inside the galactic halo\footnote{For example, this justification is explicitly mentioned in~\cite{Foster:2017hbq}, but mostly it is implicitly understood.}. 
Yet, one may desire to also obtain a better understanding from the quantum mechanical perspective. This question has already been given some attention. In Ref.~\cite{Ioannisian:2017srr} the authors investigate the power output of a (dielectric) haloscope~\cite{Caldwell:2016dcw} (see~\cite{Sikivie:1983ip} for the original (cavity) haloscope concept and~\cite{Jaeckel:2013sqa,Jaeckel:2013eha} for intermediate dish antenna ideas) using a fully quantum approach. Ref.~\cite{Beutter:2018xfx} considers the electromagnetic field amplitudes in a classical axion background. For our discussion it is noteworthy that Ref.~\cite{Ioannisian:2017srr} already comments that the specific nature of the axion quantum state can affect questions beyond the average power output. We will see exemplary cases of this in the following, while also retaining that we typically expect that the classical approximation yields good results for suitable experimental observables. 

In this note we explicitly want to consider the following questions that were raised by Dima Budker during the FIPS 2022 workshop\footnote{We would like to express our gratitude to Dima Budker for thereby starting this investigation.}. How do the oscillation frequencies that are the basis of resonant detection methods arise in a quantum mechanical setting? And how is this related to and depends on the quantum state of the setup? 
More explicitly, we could imagine a situation where axions around Earth are bound by Earth's gravitational field and are occupying the ground state of this potential well.\footnote{In many situations, we actually expect most of the axions \emph{not} to be bound to Earth, but instead to feature a distribution of velocities relative to Earth. Thereby the considered case is already special.} The question is now, what will an experiment like CASPEr~\cite{Budker:2013hfa}, designed to be sensitive to signals oscillating with a frequency equal to the axion mass, see in such a situation.

We will do so for the example of a spin precession experiment such as CASPEr~\cite{Budker:2013hfa} that aims for a detection of axions and axion-like particles coupling to the spin of nucleons. The reason for choosing this setup are three-fold. First of all, it was the context in which the question was originally brought up. Second, in spin-precession experiments the measured quantity is directly the oscillating amplitude of the spin and resulting magnetization in the directions transverse to the external magnetic field. Therefore, the oscillation is an essential ingredient in the sensitivity of the experiment. Third, the experiment can be easily translated (see Sec.~\ref{sec:transl}) into a simple model of spins and axions that is directly usable for a quantum mechanical approach.
That said, we expect that our results are more general and that our analysis can be readily translated to other types of haloscopes, e.g. ones exploiting the photon coupling.

It should be stressed that the connection between quantum and classical behavior has a long history. Famously, Jaynes and Cummings~\cite{Jaynes:1963zz} discussed the interaction of an electromagnetic field with an atom (see, e.g.~\cite{Larson:2022jvs, gerry_knight_2004} for a review and a textbook but also~\cite{JCwiki, atom-photon-interactions.ch5}).  This system is essentially the same as that of an axion interacting with a nucleon spin that we will use in the following. In this sense neither our results nor our discussion are new. Our goal is simply to explicitly check  and discuss how, for the specific case of axion experiments, the signals and in particular the different relevant time-scales expected from the standard classical calculation arise in a quantum mechanical picture. 

Our plan for the note can be outlined as follows. In Sec.~\ref{sec:transl} we introduce the quantum mechanical description of the axion field and the experimental setup. In Sec.~\ref{sec:energyeigenstate} we then discuss an initial state with a fixed number of axions and spins parallel to the magnetic field, i.e. an energy eigenstate of the system without the (weak) axion spin interaction. This will result in a situation where no oscillation of the expectation value of the magnetization with the axion mass/Larmor frequency can be observed -- seemingly a potential disaster for the sensitivity of the experiment. To remedy this we then move in Sec.~\ref{sec:superposition} to a simple superposition state, that features an oscillation of the magnetization with a frequency given by the axion mass and a magnitude close to the result of the classical calculation. Then, in Sec.~\ref{subsec:measurement} we exemplify that states that give a near classical result can be quite easily obtained by using a suitable procedure in the experiment. We take a brief look at the effects that the measurement itself has on the state and the evolution of expectation values in Appendix~\ref{app:measurement}. Instead of measuring the spin expectation value we can also consider correlators, and find that, suitable ones feature oscillations with the axion frequency also for energy eigenstates, cf. Sec.~\ref{subsec:power}. 
We further comment on the connection to the classical result in Sec.~\ref{sec:classical}. A quick summary and discussion finishes our note in Sec.~\ref{sec:conclusions}.

\section{Axion Dark Matter Experiments in Fock Space}\label{sec:transl}

To describe the axion field we use a representation in terms of creation and annihilation operators. Moreover, we assume that the field can be expressed in terms of a set of normalizable energy eigenfunctions of the Klein-Gordon equation for the axion field.\footnote{The discrete sum indicated here can always be achieved by going to a finite volume, or by considering bound states, e.g. those in the Earth's gravitational potential.}\footnote{We also neglect self interactions of the axion field.} Expressing the field operator in the  interaction picture we have\footnote{We choose the $\phi_{n}(x)$ to be real and normalize $\int_{x} \phi_{n}\phi_{m}=\delta_{nm}$.},
\begin{equation}
\label{eq:modeexpansion}
\phi(t,x)=\sum_{n}\frac{1}{\sqrt{2E_n}}\left[\phi_{n}(x)a_{n}\exp(-iE_{n}t)+\phi_{n}(x)a^{\dagger}_{n}\exp(+iE_{n}t)\right].
\end{equation}

The Hamiltonian can then be written as,
\begin{eqnarray}
H&=&
H_{ax}+H_{int}+H_{exp}
\\\nonumber
&=&\int d^{3}x \,\frac{1}{2}\left[(\dot{\phi}(x))^2+(\nabla\phi(x))^2+m^{2}_{a}\phi^2\right]+H_{int}+H_{exp}
\\\nonumber
&=&\sum_{n}E_{n}\left(a^{\dagger}_{n}a_{n}+\frac{1}{2}\right)+H_{int}+H_{exp}.
\end{eqnarray}
Here, $E_{n}$ are the energy eigenvalues of the eigenstates described by the wave functions $\phi_{n}(x)$. $H_{int}$ is the interaction term, to which we turn below, and $H_{exp}$ is the experimental Hamiltonian in absence of axions.

We stress that the $E_{n}$ correspond to the full relativistic energy of the state $n$, including the axion rest mass. Indeed, for typical situations, such as axions trapped in the gravitational potential of Earth, the binding energy is much smaller than the mass and we can approximate,
\begin{equation}
E_{n}\approx m_{a}+{\rm small\,\,correction}.
\end{equation}
As a concrete example situation we chose an experiment like CASPEr, that is targeting interactions of the axion with nucleonic spins.
For simplicity we treat the latter as a set of non-self-interacting, non-moving spin 1/2 states located close to $x=0$. The spins are interacting with a magnetic field in the $z$-direction, generating an energy splitting set by the Larmor frequency.
The corresponding Hamiltonian is,
\begin{equation}
H_{exp}=-\frac{\omega_{L}}{2}\sum_{i=1}^{\Nspin}\sum_{s,s' = \uparrow, \downarrow}\left[b^{\dagger}_s \left( \sigma_z\right)_{s s'}b_{s'}\right]_i,
\end{equation}
where $b^{\dagger}_{s,i}$ and $b_{s,i}$ denote the creation and annihilation operators for the spin state $s$ of the $i$-th nucleon. Moreover, $\vec{\sigma}=(\sigma_{x},\sigma_{y},\sigma_{z})$ are the Pauli matrices.

Finally, the relevant bit to describe the interaction with the experiment is, of course, the interaction Hamiltonian $H_{int}$.
For concreteness let us consider an interaction with an axion induced electric dipole operator, cf.~\cite{Graham:2013gfa},
\begin{eqnarray}
H_{int}=-\int d^{3}x {\mathcal{L}}_{int}=\int d^{3}x \frac{i}{2}g_{d}\phi(x)F_{\mu\nu}\bar{\psi}\sigma^{\mu\nu}\gamma^{5}\psi.
\end{eqnarray}
We implement the electromagnetic field as an external constant electric field $\vec{E}$ and, as already mentioned, the matter fields $\psi$ as a set of $\Nspin$ non-moving and non-interacting spins located (close) to $x=0$,
\begin{equation}
\psi\approx\sum_{i=1}^{N} \sum_{s=\uparrow,\downarrow} \sqrt{\delta(x)}\exp(-i\epsilon_{0}t-i\epsilon_{s}t) u_{s} b_{s,i}
\end{equation}
where, as befitting a non-relativistic situation, we neglect the antiparticle contributions. Accordingly, the matter field operator only contains the creation operators $b_{s,i}$ for the spin up and down particle state in the chosen location.
$\epsilon_{0}$ denotes the spin-independent energy of the chosen state (which will drop out in the following) and $\epsilon_{s}$ the spin-dependent one.  
The spinors and spin-state energies are given by
\begin{eqnarray}
u_{\uparrow}&=&(1,0,0,0)^{T},\qquad \epsilon_{\uparrow}=-\frac{1}{2}\omega_{L},
\\\nonumber
u_{\downarrow}&=&(0,1,0,0)^{T},\qquad \epsilon_{\downarrow}=\frac{1}{2}\omega_{L}.
\end{eqnarray}

Using this ansatz simplifies the interaction term (see, e.g.~\cite{Bernreuther:1990jx})
\begin{equation}
\frac{i}{2}F_{\mu\nu}\bar{\psi}\sigma^{\mu\nu}\gamma^{5}\psi=2\vec{E}
\hat{\vec{S}}\delta(x).
\end{equation}
Here, $\hat{\vec{S}}$ is the spin operator for the full sets of spins that can be expressed as \cite{MAN_KO_1994}
\begin{equation}
\hat{\vec{S}}(t)=\frac{1}{2}\sum_{i=1}^{\Nspin}\sum_{s,u=\uparrow,\downarrow}b^{\dagger}_{s,i}(\vec{\sigma})_{su}b_{u,i}\exp(i(\epsilon_{s}-\epsilon_{u})t).
\end{equation}

This becomes even more concrete if we specialize to the experimentally relevant case where the electric field points in the $x$-direction. In this case we only need this component of the spin that can be conveniently expressed in terms of the spin creation and annihilation operators as follows,
\begin{equation}
\hat{S}_{x}(t)=\frac{1}{2}\sum_{i=1}^{\Nspin}\sigma_{x,i}(t)=\frac{1}{2}\sum_{i=1}^{\Nspin}\left[b^{\dagger}_{\uparrow,i}b_{\downarrow,i}\exp(-i\omega_{L}t)+b^{\dagger}_{\downarrow,i}b_{\uparrow,i}\exp(+i\omega_{L}t)\right].
\end{equation}
Putting everything together we have,
\begin{eqnarray}
\label{eq:intham}
H_{int}(t)&=&g_{d} E_{x}(0)\sum_{n}\frac{1}{\sqrt{2E_{n}}}\phi_{n}(0)\sum_{i=1}^{\Nspin}\bigg[a_{n}b^{\dagger}_{\downarrow,i}b_{\uparrow,i}\exp(-i(E_{n}-\omega_{L})t)
\\\nonumber
&&\qquad\qquad\qquad\qquad\qquad\qquad\qquad\quad+a^{\dagger}_{n}b^{\dagger}_{\uparrow,i}b_{\downarrow,i}\exp(+i(E_{n}-\omega_{L})t)\bigg].
\end{eqnarray}
The time dependence arises from the time dependence of the respective states/operators and effectively also implements energy conservation. Roughly speaking, integrating over an (infinite) time interval we would get a $\delta$-function in energy. This is also the reason for dropping terms with $\sim \exp(\pm i(E_{n}+\omega_{L})$ that are oscillating more quickly. This is the so-called rotating wave approximation (RWA) (cf., e.g.~\cite{RWwiki, atom-photon-interactions.ch5}).\footnote{This is indeed an approximation and we will briefly comment on it again, when comparing to the classical results in Sec.~\ref{sec:classical}.} 

Let us note at this point that the simplified model obtained above is essentially a Jaynes-Cummings model~\cite{Jaynes:1963zz}. As already mentioned in the introduction this model has been used to extensively study the relation between classical and quantum physics. Our aim here is only to explicitly apply it to the case of axion detection experiments.

We can now further simplify to a situation where all axions are in the same spatial state, i.e. only one energy state $E$ with wave function $\varphi$ is occupied. Note, that this does not yet fix the quantum state of the axion field, because this spatial state can be occupied by different numbers of axions (they are bosons after all). Moreover, let us choose the magnetic field such that the Larmor frequency matches the energy of the single axion $\omega_{L}=E\approx m_{a}$.
In this case the interaction simplifies to, 
\begin{eqnarray}
H_{int}=\frac{g_{d} E_{x}(0)}{\sqrt{2E_{n}}}\varphi(0)\sum_{i=1}^{\Nspin}\bigg[a b^{\dagger}_{\downarrow,i}b_{\uparrow,i}
+a^{\dagger}b^{\dagger}_{\uparrow,i}b_{\downarrow,i}\bigg].
\end{eqnarray}

There is no explicit time-dependence (other than from a possibly time-dependent electric field).
Energy conservation is now realized in the following sense. $b^{\dagger}_{\downarrow,i}b_{\uparrow,i}$ takes a single spin up state and flips it into a down state. This costs an energy $\omega_{L}$. This is compensated by the operator $a$ ``destroying'' one axion of energy $E\approx m_{a}\approx\omega_{L}$. Hence energy is conserved.
Indeed, let us now look at a state in Fock space that contains $\Naxion$ axions and where $\Nup$ spins are up and $\Ndown$ spins are down,
\begin{equation}
|\Naxion,\Nup,\Ndown\rangle.
\end{equation}
Then the effect the interaction Hamiltonian is structurally as follows,
\begin{equation}
H_{int}|\Naxion,\Nup,\Ndown\rangle= A |\Naxion-1,\Nup-1,\Ndown+1\rangle+B |\Naxion+1,\Nup+1,\Ndown-1\rangle,
\end{equation}
with some constants $A,B$.
Crucially it is really the destruction of an axion that provides the energy for the spin flip. Therefore, the full relativistic energy, including the rest mass energy, is available for this process.

Let us briefly compare this to the case of photon emission from an excited atom (as indeed described by the Jaynes Cummings Hamiltonian~\cite{Jaynes:1963zz}). There, the number of electrons is unchanged\footnote{This is already required by lepton number conservation.}. Instead the relevant operators would destroy the excited electron state and create a lower energy electron state. Importantly, while the state of the electrons in the atom changes, the total number of electrons is unchanged. Hence, the energy available for the creation of the photon is only the difference in the binding energies. Indeed, if we want a better analogy with our above interaction Hamiltonian, we can compare the electron states to the spin states and the axion to the photon. The total number of spins is conserved and the energy difference is only given by the binding energy in the magnetic field, similar to the case of the electrons in the atom. In contrast neither axion nor photon number are conserved and their full energy is required/available in the process.

\section{Energy eigenstate}\label{sec:energyeigenstate}

Let us now turn to a concrete example of an initial state and study its time evolution.

The naively simplest initial state is an energy eigenstate of $H_{ax}+H_{exp}$, i.e. an energy eigenstate in the absence of the dipole interaction.
Considering states in the absence of interactions makes sense because, for example, we could switch off the electric field $E_{x}$ before starting the experiment. Moreover, the axion interaction with the experiment is always very weak and we can therefore treat it as a small perturbation.

More concretely, we consider a situation where we start with an axion number eigenstate with $\Naxion$ axions and all spins parallel to the magnetic field $\Nup=\Nspin$ and $\Ndown=0$,
\begin{equation}
\label{eq:energyeigenstate}
|\Naxion,\Nspin,0\rangle.
\end{equation}

In absence of interactions there is a set of energy eigenstates degenerate with the above state,
\begin{equation}
\label{eq:subspace}
|\Naxion-\Ndown,\Nspin-\Ndown,\Ndown\rangle,\qquad N_{\downarrow}=0,\ldots,N_{s}.
\end{equation}
Therefore, following the logic of degenerate perturbation theory, in the presence of weak interactions we can now restrict our discussion to this degenerate subspace.

\subsection{$\Nspin=1$}
The simplest case is a single spin, i.e. $\Nspin=1$.
In this case we have only two degenerate states,
\begin{equation}
|\Naxion,1,0\rangle,\quad |\Naxion-1,0,1\rangle.
\end{equation}
In this subspace our interaction Hamiltonian is given by,
\begin{equation}
H_{\rm int}= \oc \left(\begin{array}{cc}
0&\sqrt{\Naxion}
\\
\sqrt{\Naxion} & 0
\end{array}\right).
\end{equation}

Before solving the time evolution we remark that the effective coupling frequency,
\begin{equation}
\label{eq:omegac}
    \oc \sqrt{\Naxion}=\frac{g_{d} E_{x}(0)}{\sqrt{2E_{n}}}\varphi(0)\sqrt{\Naxion}=\frac{g_{d} E_{x}(0)}{\sqrt{2E_{n}}}\sqrt{2\rho_{a}/m_{a}},
\end{equation}
is exactly the combination that appears also in the classical field theory calculation (more details in Sec.~\ref{sec:classical}).
Here, $\rho_{a}$ is the axion energy density and accordingly $\rho_{a}/m_{a}$ is the number density.
Moreover,  we have used that $\varphi^2(0)$ gives the probability density for an axion to be located in the vicinity of $x=0$, if there is only a single axion. This increases by the number of axions, $N_a$, occupying the same state.

Starting from the initial state $(1,0)^{T}$ we can now obtain the time evolution.
This is most easily done in the interaction picture, where we factor out the oscillation $\sim E=\omega_{L}$ from the non-interacting parts $H_{ax}$ and $H_{exp}$.
The result is
\begin{equation}
\Psi(t)_I = \left(\cos(\sqrt{\Naxion}\omega_ct),-i\sin(\sqrt{\Naxion}\omega_c t)\right)^T.
\label{eq:res-Ns=1}
\end{equation}

We can already see that the time-scale of the evolution is only set by the coupling frequency $\oc$, that is suppressed by the small coupling $g_{d}$.
The axion energy/mass and the Larmor frequency do not explicitly appear.

\begin{figure}
\centering
\includegraphics[width=.7\textwidth]{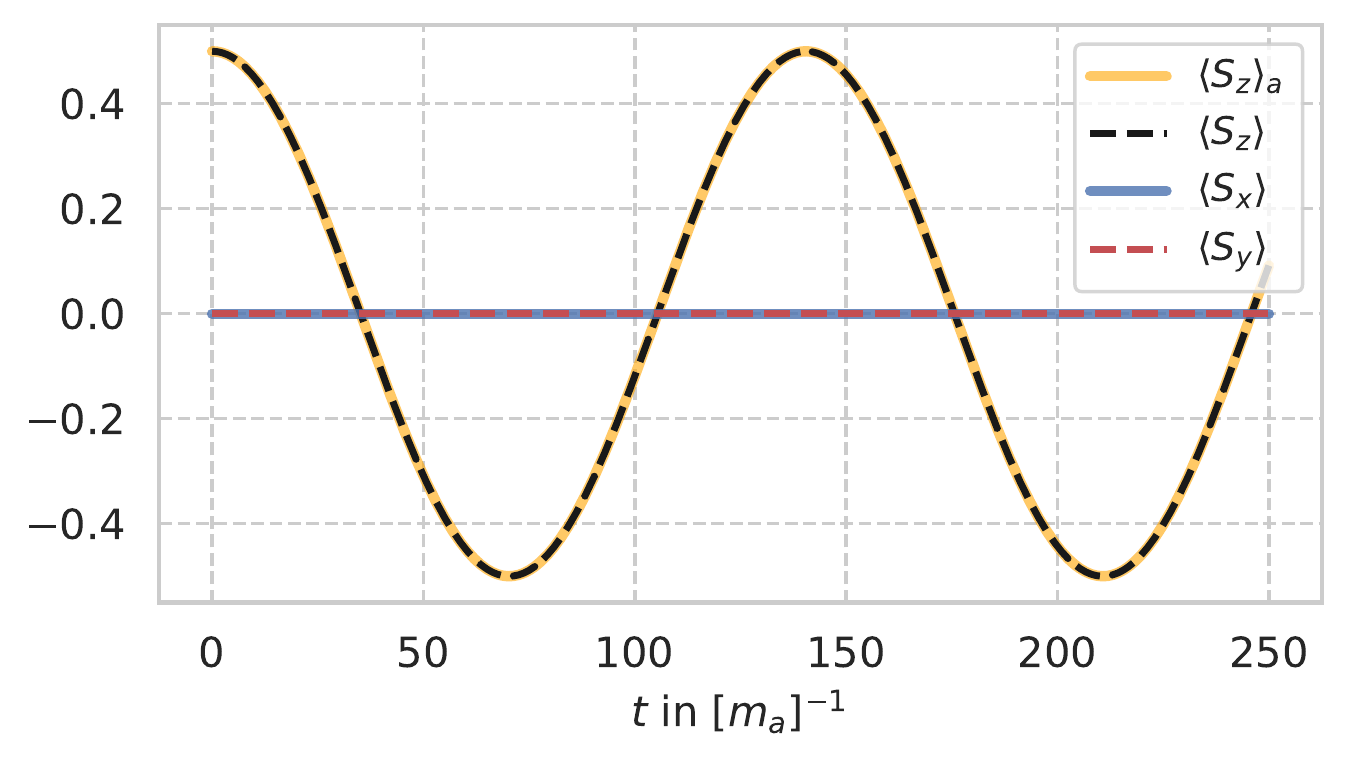}
\caption{Time evolution of the expectation values of the spin operator for the state $|5,1,0\ket$ with coupling frequency $\omega_c= .01 \, [m_a]$, $m_a = 1 \, [m_a]$. $\langle S_z\rangle_a$ denotes the analytical result for the $z$-direction given by equation \eqref{eq:res-Ns=1}. The other results were obtained numerically using the Qutip library \cite{JOHANSSON20121760,JOHANSSON20131234}. Everything is kept in units of the axion mass, such that the result is invariant under rescalings. We see an oscillation only for the expectation value $\langle S_z \rangle$. Also note that the oscillation frequency in this signal is given by twice the coupling frequency $\sqrt{N_{a}}\omega_c$ as the expectation value depends on $\cos^2 (\sqrt{N_{a}}\omega_ct) - \sin^2(\sqrt{N_{a}}\omega_ct)$. 
}
\label{fig:singlespin}
\end{figure}

One may wonder whether this is an artifact of the interaction picture. However, the change to the interaction picture is an oscillation that is equal for all relevant states. Therefore, it does not affect the expectation value of observables such as,
\begin{equation}
\langle S_{z}\rangle,\quad \langle S_{y}\rangle ,
\end{equation}
calculated between states within the degenerate subspace . This is also shown in Fig.~\ref{fig:singlespin} for an example with $N_a=5$.

The absence of an oscillation with the Larmor frequency is perhaps not too surprising. We have started from an energy eigenstate. Energy eigenstates are stationary (i.e. expectation values involving them do not depend on time). 
The only real time dependence therefore arises from the fact hat the interaction Hamiltonian slightly changes the energy eigenstates and eigenvalues. The time-evolution then arises due to this small splitting in frequencies $\sim \oc$.

However, this may also seem slightly worrying because the sensitive observable measured in CASPEr is supposed to be the transverse magnetization rotating with the Larmor frequency; effectively this should correspond to $\langle S_{y}\rangle(t)$.
As we discuss in detail in the next section~\ref{sec:superposition} the absence of this is an artifact of the special initial state we have taken, as well as the specific quantity that we have chosen to measure. 
For now, let us remark that in an energy eigenstate the expectation value of the field operator which, in the classical approximation we would want to identify with the classical field value, vanishes.\footnote{The (non-vanishing) energy is essentially stored in the expectation value of the \emph{square} of the field value.}\footnote{The vanishing expectation value can be understood as follows. Energy eigenstates are stationary, i.e. the phases $\exp(-iEt)$ from their time evolution drop out when calculating expectation values. Therefore, the expectation value of the field operator is time independent. As we do not have a situation with spontaneous symmetry breaking, the constant value must indeed be zero.}
In this sense one could say there is no proper classical field to drive the expectation values of $S_{x,y}$.

\subsection{$\Nspin>1$}

Let us now briefly also discuss the situation with more than one spin.

We consider the, in absence of interactions, degenerate subspace, Eq.~\eqref{eq:subspace}, spanned by \linebreak \mbox{$|\Naxion-\Ndown,\Nspin-\Ndown,\Ndown\ket$,} where $0\leq\Ndown \leq \Nspin$ . The interaction Hamiltonian in this basis is given by
\begin{eqnarray}
 \left( H_{int} \right)_{ij} &=& \bra \Naxion-i+1, \Nspin-i+1, i-1 | H_{int} | \Naxion-j+1 , \Nspin-j+1,j-1 \ket \nonumber\\
 &=& \omega_c(\delta_{j,i+1}+\delta_{j+1,i})\sqrt{(\min(i,j))(\Nspin-\min(i,j)-1)}\sqrt{\Naxion-\min(i,j)-1}.
\end{eqnarray}
Explicitly, the matrix representation of the interaction Hamiltonian is schematically given by
\begin{equation}
    H_{int} = \omega_c\begin{pmatrix}0 & \sqrt{\Naxion \Nspin} & 0 &\dots &\dots& 0 \\ \sqrt{\Naxion \Nspin} & 0 &\sqrt{2(\Nspin-1)(\Naxion-1)}&0& \dots & 0\\ \vdots& \ddots& \ddots & \ddots&\vdots&\vdots \end{pmatrix}\,.
    \label{eq:Hint}
\end{equation}

Let us briefly understand how this matrix is obtained. Structurally it is a symmetric band matrix with non-vanishing entries only on the first sub- and superdiagonal. This can be understood from the interaction Hamiltonian which always acts on any energy eigenstate by flipping one spin and destroying or creating one axion to either supply the needed energy, or store the supplied energy. 

To obtain the entries of this matrix we can use the normalisation of the degenerate energy eigenstates, and the combinatorics arising when applying the interaction Hamiltonian. A state of the form $| \Naxion-\Ndown, \Nspin-\Ndown,\Ndown\ket$ is the tensor product of an axion number state with a superposition of spin product states that have the correct number of up- and down-spins. To ensure the orthonormality of such states, one needs to account for the number of possible product states that contain $\Ndown$ down- and $\Nspin-\Ndown$ up-states. This is the same as counting the number of subsets with $\Ndown$ elements within a set of $\Nspin$ elements. Therefore, the normalisation given by 
\begin{equation}
    | \Naxion-\Ndown, \Nspin-\Ndown,\Ndown\ket = {\Nspin \choose \Ndown}^{-1/2} \left(|\Naxion-\Ndown\ket \otimes \sum_i |S(\Ndown)_i\ket \right).
\end{equation}
Here, the states $|S(\Ndown)_i\ket$ are the product states with the correct number of up- and down-spins. When applying the interaction Hamiltonian, one has to account for the number of possible flips that can occur when projecting onto another state. Consider the part of the interaction that projects the state $| \Naxion-(\Ndown+1),\Nspin-(\Ndown+1),\Ndown+1\ket$ onto  $| \Naxion-\Ndown,\Nspin-\Ndown,\Ndown\ket$. For every product state, there are $\Ndown+1$ possible flips to be carried out. But since the resulting state then has less spins in the down position, there will be redundancy present in the resulting superposition. This redundancy shows up in the interaction as the number of product states within the initial state,
\begin{eqnarray}
    ( H_{int})_{\Ndown-1,\Ndown} &=& \sqrt{\Naxion-\Ndown}\left({\Nspin \choose \Ndown}{\Nspin \choose \Ndown+1}\right)^{-1/2}\cdot (\Ndown +1 ){\Nspin \choose \Ndown+1}\nonumber \\[1em] &=& \sqrt{(\Naxion-\Ndown)(\Ndown+1)(\Nspin-\Ndown)}.
\end{eqnarray}
The first two factors besides the axion number arise from the normalisation, whereas the factors after are from he application of the interaction as discussed above. The elements on the subdiagonal can be obtained analogously.

\begin{figure}
\centering
\includegraphics[width=.7\textwidth]{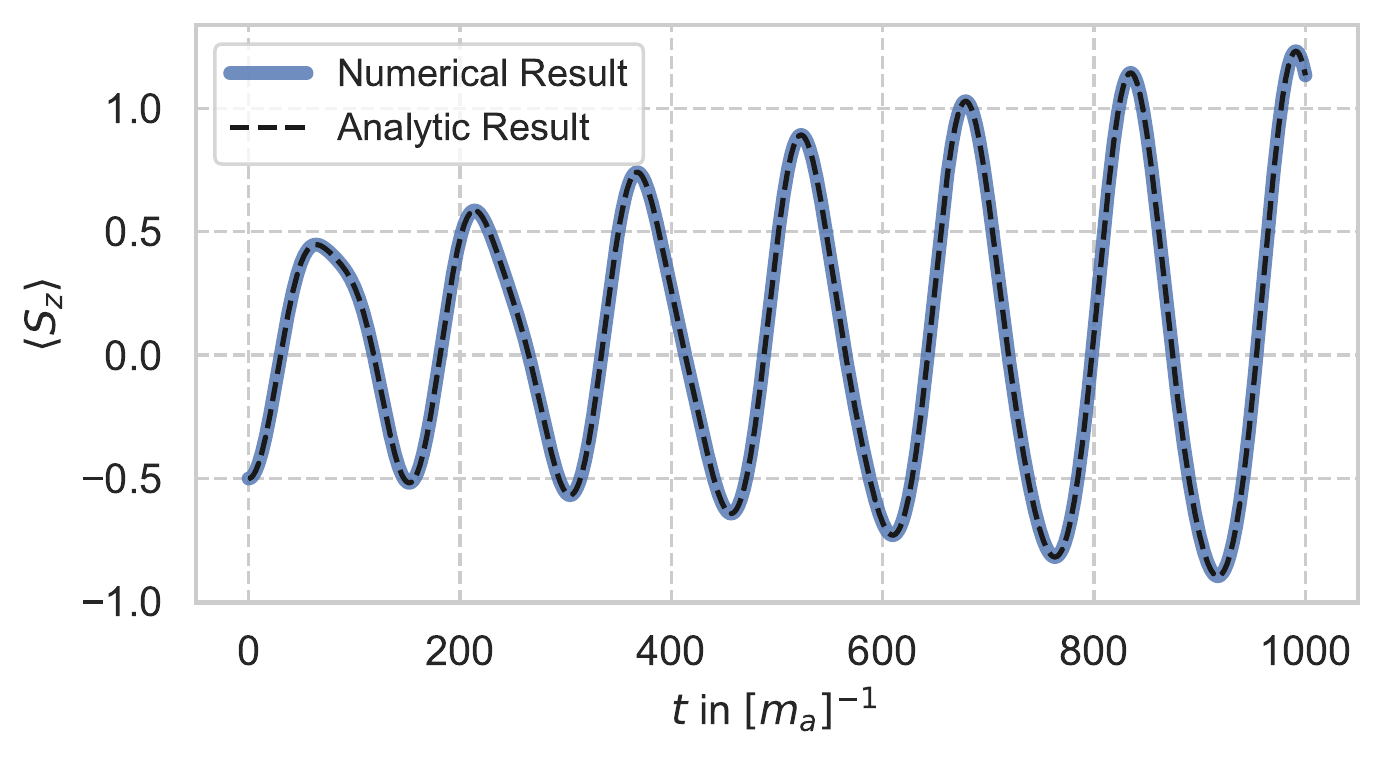}
\caption{Expectation values $\langle S_z \rangle$ for the initial energy eigenstate $|5,1,2\ket$. The axion mass in this computation was taken as $m_a = 1 \, [m_a]$, and the coupling frequency given via $\oc =0.01\, [m_a]$.
As a check we show a numerical result obtained via Qutip~\cite{JOHANSSON20121760,JOHANSSON20131234} (black, dashed), and  an analytical one (obtained with Sympy~\cite{10.7717/peerj-cs.103})  from the interaction Hamiltonian in Eq.~\eqref{eq:Hint} (blue, solid).  }
\label{fig:nonsimpleexample}
\end{figure}

\bigskip

Again we can now calculate the evolution of an original energy eigenstate using this interaction Hamiltonian. An example is shown in Fig.~\ref{fig:nonsimpleexample}.  However, as before, we find that for $\langle S_{x}\rangle$ and $\langle S_{y}\rangle$ no oscillation with the Larmor frequency appears. Indeed they both continue to vanish. Without going into a detailed calculation this can be understood from the fact that, again, the energy splittings in the degenerate sub-space are all proportional to the small\footnote{Compared to the Larmor frequency $\omega_{L}$.} frequency $\omega_{c}$. Hence, no oscillation with the Larmor frequency appears. More technically, we can also see that application of the spin operators in the $x$- or $y$-direction raises or lowers the spin in each state by one unit. If we leave the axion state unchanged (as appropriate for a pure spin measurement) the resulting state is then a combination of states whose energy differs from the original energy by one unit of the Larmor frequency. These states are therefore orthogonal to the original state and the expectation value vanishes. That said it makes apparent that this can be remedied if the starting state already contains a combination of energy eigenstates with energies differing by one unit of $\omega_L$. This is what we will look at next.

\section{The appearance of oscillations with the axion mass and suitable measurement procedures}\label{sec:superposition}

In the previous section we have seen that for particle number eigenstates we do not find an expectation value of the spin oscillating with the Larmor frequency that is usually the naive observable in the experiment.

A simple argument to avoid this is that one would not really expect the axion to be in a particle number eigenstate. Indeed it is usually argued that a coherent or Glauber state~\cite{Glauber:1963tx} is a good assumption for the axion field~\cite{Ioannisian:2017srr}.
However, one might worry that, in particular for axions trapped in some potential, there exist relaxation processes that may bring us closer to a particle number state. In consequence one may fear that the signal is significantly reduced or even absent.

However, we think that in practice this is not a problem.
First, already small modifications away from the pure axion number state result in an oscillating signal with an amplitude that is close to the classical expectation. We show this by giving an example in the next subsection. Therefore, the initial state would have to be rather close to the number eigenstate in order to have a suppression of the signal by orders of magnitude -- a situation we think is unlikely as we do not see any strong processes forcing the system into a number eigenstate.
Second, even if the system is in an exact number eigenstate, one can adapt the experimental procedure in such a way that the system is modified away from an energy eigenstate into a state where oscillations with $\omega_{L}=E\approx m_{a}$ will occur. Third, we discuss that measuring a suitable quantity, in particular a correlator related to the power spectrum, can avoid the problem of the vanishing expectation value altogether.
Our approach is to give some explicit, but rather crude, examples. A more realistic study of the measurement procedures and their effects as they can be implemented in CASPEr would be desirable but is beyond the scope of the present note (that said, we make some additional comments on the effects of measurements in Appendix~\ref{app:measurement}).

\subsection{A simple initial state oscillating with frequency $\omega_{L}=E\approx m_{a}$}
\label{subsec:simpleexample}

\begin{figure}
    \centering
    \includegraphics[width=.7\textwidth]{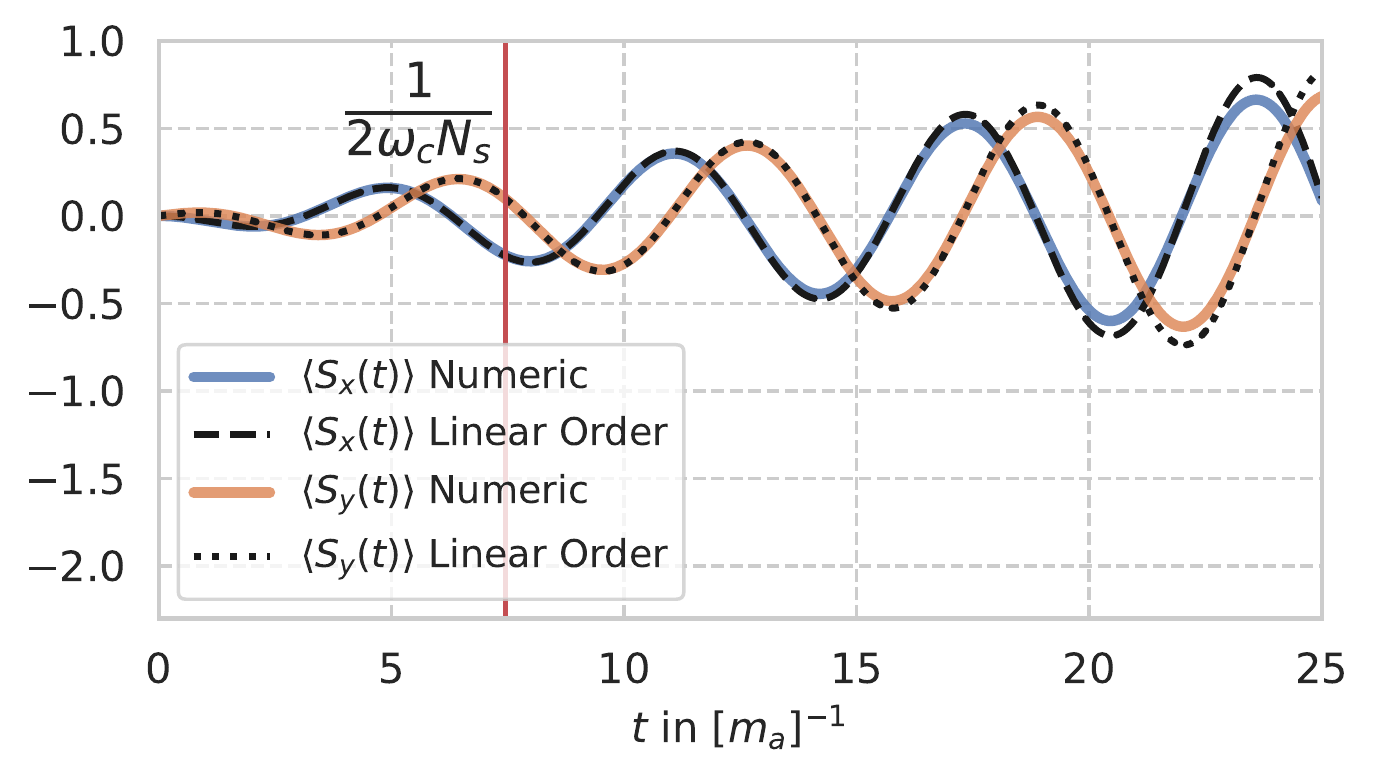}
    \caption{Time evolution of the spin expectation values in the directions transverse to the magnetic field. The chosen initial state is $\frac{1}{\sqrt{2}}(|5\ket + |4\ket)|3,0\ket$.  The coupling frequency and axion mass/Larmor frequency are given by $\omega_c = .01\, [m_a]$ and $m_a =1 \, [m_a]$. As expected we observe an oscillating signal.  Also as expected we find good agreement between the linear order result Eq.~\eqref{eq:simplestateresult} and the numerical result obtained with Qutip~\cite{JOHANSSON20121760,JOHANSSON20131234}.}
    \label{fig:osci}
\end{figure}

One of the simplest non-axion particle number eigenstates is the following,
\begin{equation}
\label{eq:superposstate}
    | \Psi_0 \ket = \frac{1}{\sqrt{2}}\left( |\Naxion\ket + |\Naxion-1\ket\right) \otimes|\Nspin\uparrow,0\downarrow\ket.
\end{equation}
As we can see in Fig.~\ref{fig:osci} this already produces the desired oscillation with $\omega_L$.

\bigskip

To convince ourselves that this is not a special feature of this particular state, let us approximately analyse the signal for a more general initial state $| \Psi_0 \ket$. In a first order approximation in the interacting Hamiltonia, i.e. in the small frequency $\omega_c$ we can obtain the linearised time evolution, exploiting the smallness of the effective coupling frequency. 

To this end, we consider the time evolution of the proposed initial state in the interaction picture, and linearise as follows
\begin{equation}
\label{eq:examplestate}
    |\Psi_I(t)\ket = e^{iH_0t}e^{-iHt}|\Psi_0 \ket =(1 -i(H-H_0)t + \mathcal{O}(t^2))|\Psi_0 \ket.
\end{equation}
The time evolution of the spin operators (acting on the $i$th spin) in the interaction picture is given by,
\begin{eqnarray}
    \frac{1}{2}e^{iH_0t}\sigma_{i,x}e^{-iH_0t} &=& \frac{1}{2}\left[e^{im_at}\sigma^{-}_{i} + e^{-im_at}\sigma^{+}_{i}\right],\\\nonumber
    \frac{1}{2}e^{iH_0t}\sigma_{i,y}e^{-iH_0t} &=& \frac{1}{2}\left[ie^{im_a t}\sigma^{-}_{i} -i e^{-im_at}\sigma^{+}_{i}\right],
\end{eqnarray}
where, we simplified our notation for the spin-flip operators to be $b^{\dagger}_{i,\downarrow}b_{i,\uparrow} = \sigma_i^-$ and $b^{\dagger}_{i,\uparrow}b_{i,\downarrow} = \sigma_i^+$.

Putting everything together, we obtain the expectation values
\begin{eqnarray}
 \langle S_x \rangle\!\! &=& \!\!\sum_{i=1}^{N_{s}}\frac{1}{2}\bra \Psi_0| (1+iH_{\text{int}}t) (e^{im_at}\sigma^{-}_{i}
+ e^{-im_at}\sigma^{+}_{i} )(1-iH_{\text{int}}t) | \Psi_0 \ket
 \\\nonumber
 \!\!&\approx&\!\! \sum_{i=1}^{N_{s}}\bra \Psi_0|\frac{\omega_c}{2} \bigg[ e^{im_a t}(\sigma^{-}_{i} + ia^{\dagger}\sigma^{+}_{i} \sigma^{-}_{i} t + ia \sigma^{-}_{i} \sigma^{-}_{i}t ) 
  \\\nonumber 
 &&\qquad\qquad\qquad\qquad\qquad\qquad\qquad
 +e^{-im_a t} (\sigma^{+}_{i} + ia \sigma^{+}_{i} \sigma^{-}_{i} t + ia^{\dagger}\sigma^{+}_{i} \sigma^{+}_{i} t) \bigg] | \Psi_0 \ket \nonumber \\
   \langle S_y \rangle\!\! &=&\!\! \sum_{i=1}^{N_{s}}\frac{1}{2}\bra \Psi_0| (1+iH_{\text{int}}t) i(e^{im_at}\sigma^{-}_{i} - e^{-im_at}\sigma^{+}_{i} )(1-iH_{\text{int}}t) | \Psi_0 \ket
   \\\nonumber
  \!\! &\approx&\!\! \sum_{i=1}^{N_{s}}\bra \Psi_0|\frac{i\omega_c}{2}\bigg[ e^{im_a t}(\sigma^{-}_{i} + ia^{\dagger}\sigma^{+}_{i} \sigma^{-}{i} t + ia \sigma^{-}_{i} \sigma^{-}_{i}t ) 
   \\\nonumber 
 &&\qquad\qquad\qquad\qquad\qquad\qquad\qquad-e^{-im_a t} (\sigma^{+}_{i} + ia \sigma^{+}_{i} \sigma^{-}_{i} t + ia^{\dagger}\sigma^{+}_{i} \sigma^{+}_{i} t) \bigg]| \Psi_0 \ket
   .
\end{eqnarray}
Because there is always either a single spin change operator or an axion number changing operator, we see that for any oscillation or indeed any time variation to occur in linear order, the state $|\Psi_0\ket$ has to be a superposition: The states involved need to exhibit some mismatch in their spin or axion number for the expectation value not to be projected out to zero.

This can be explicitly seen for the example state Eq.~\eqref{eq:superposstate},
\begin{eqnarray}
\label{eq:simplestateresult}
    \langle S_x \rangle &=& \sum_{i=1}^{N_{s}}i \omega_{c} t\left(\frac{e^{im_a t}}{2}\bra \Psi_0 |a^{\dagger} \sigma^{+}_{i} \sigma^{-}_{i} |\Psi_0 \ket - \frac{e^{-im_a t}}{2}\bra\Psi_0|  \sigma^{+}_{i} \sigma^{-}_{i} a|\Psi_0 \ket\right) \nonumber \\
    &=& \frac{\omega_c \sqrt{\Naxion}\Nspin t}{2} \sin (m_a t),\nonumber\\
    \langle S_y \rangle &=& \sum_{i=1}^{N_{s}}- \omega_c t\left(\frac{e^{im_a t}}{2}\bra \Psi_0 |a^{\dagger} \sigma^{+}_{i} \sigma^{-}_{i} |\Psi_0 \ket + \frac{e^{-i\omega_L t}}{2}\bra\Psi_0|  \sigma^{+}_{i} \sigma^{-}_{i} a|\Psi_0 \ket\right) \nonumber \\
    &=&  -\frac{\omega_c \sqrt{\Naxion}\Nspin t}{2} \cos (m_a t)\,.\nonumber
\end{eqnarray}
The only parts contributing to the above expectation values, calculated at linear order in time, are states with axion numbers differing by $1$. 

In turn we can see from this that already for a simple modification of an energy eigenstate, i.e. one that contains states with axion numbers differing by 1, we  will get oscillations  of the transverse magnetisation with the Larmor frequency. 

Note that the amplitude of the oscillations in Eq.~\eqref{eq:simplestateresult} exhibits the scaling $\sim \sqrt{N_{a}}N_{s}$ that we expect in the classical approximation. We will return to a more detailed comparison between the classical and quantum results in Sec.~\ref{sec:classical}.

\subsection{Using the experiment to change the state away from an axion number eigenstate}\label{subsec:measurement}

In the previous subsection we have seen that already a relatively modest deformation away from the energy eigenstate allows us to have an oscillating spin expectation value.

Let us now see that such a modification can already be obtained by a operating the experiment in a suitable way.

For simplicity let us focus on the simplest case of only one spin for the calculation, but we comment on the case with more spins when appropriate. Using the initial state $|\Naxion,1,0\rangle$ we can simply use the result for the time evolved state, Eq.~\eqref{eq:res-Ns=1}, explicitly writing the state vectors,
\begin{equation}
    |\Psi(t) \ket = \cos\left(\sqrt{N_{a}}\omega_{c}t\right) |\Naxion \rangle \otimes |\uparrow \rangle +i \sin\left(\sqrt{N_{a}}\omega_{c}t\right)|\Naxion-1\rangle \otimes |\downarrow \rangle \,.
\end{equation}

Now we can imagine that after some time $t_{stop}$ we switch off the experiment. For example, we could do so by switching off the electric field but we will discuss more efficient ways to do so momentarily.

After switching off the experiment the state would continue to evolve with the ``free'' Hamiltonian.
However, if we wait for a much longer time we would expect that also spin interactions leading to the spin relaxation play a role (they are not included in our simple Hamiltonian). In particular, if we wait longer than the spin relaxation time, the spins return to their position aligned with the magnetic field. The axion field, being much more weakly coupled, is not affected by this.
\begin{eqnarray}
\nonumber
|\Psi(t) \ket \!\!\!&=&\!\!\!\cos\left(\sqrt{N_{a}}\omega_{c}t_{stop}\right)\exp(i\delta_{\uparrow})|N_{a}\rangle \otimes |\uparrow \rangle+\sin\left(\sqrt{N_{a}}\omega_{c}t_{stop}\right)\exp(i\delta_{\downarrow})|N_{a}-1\rangle \otimes |\uparrow\rangle
\\
\!\!\!&=&\!\!\!\left[\cos\left(\sqrt{N_{a}}\omega_{c}t_{stop}\right)\exp(i\delta_{\uparrow})|N_{a}\rangle +\sin\left(\sqrt{N_{a}}\omega_{c}t_{stop}\right)\exp(i\delta_{\downarrow})|N_{a}-1\rangle\right]  \otimes |\uparrow\rangle.
\label{eq:relaxed}
\end{eqnarray}
Here, the phases $\delta_{\uparrow},\,\delta_{\downarrow}$ between the two different eigenstates depends on the relaxation process.

Nevertheless, the state Eq.~\eqref{eq:relaxed} is a combination of states with different energy, similar to Eq.~\eqref{eq:superposstate}. 
Therefore, if we now switch the experiment on again, we will typically find an oscillating spin transverse to the magnetic field. 

\begin{figure}
    \centering
   \includegraphics[width=.8\textwidth]{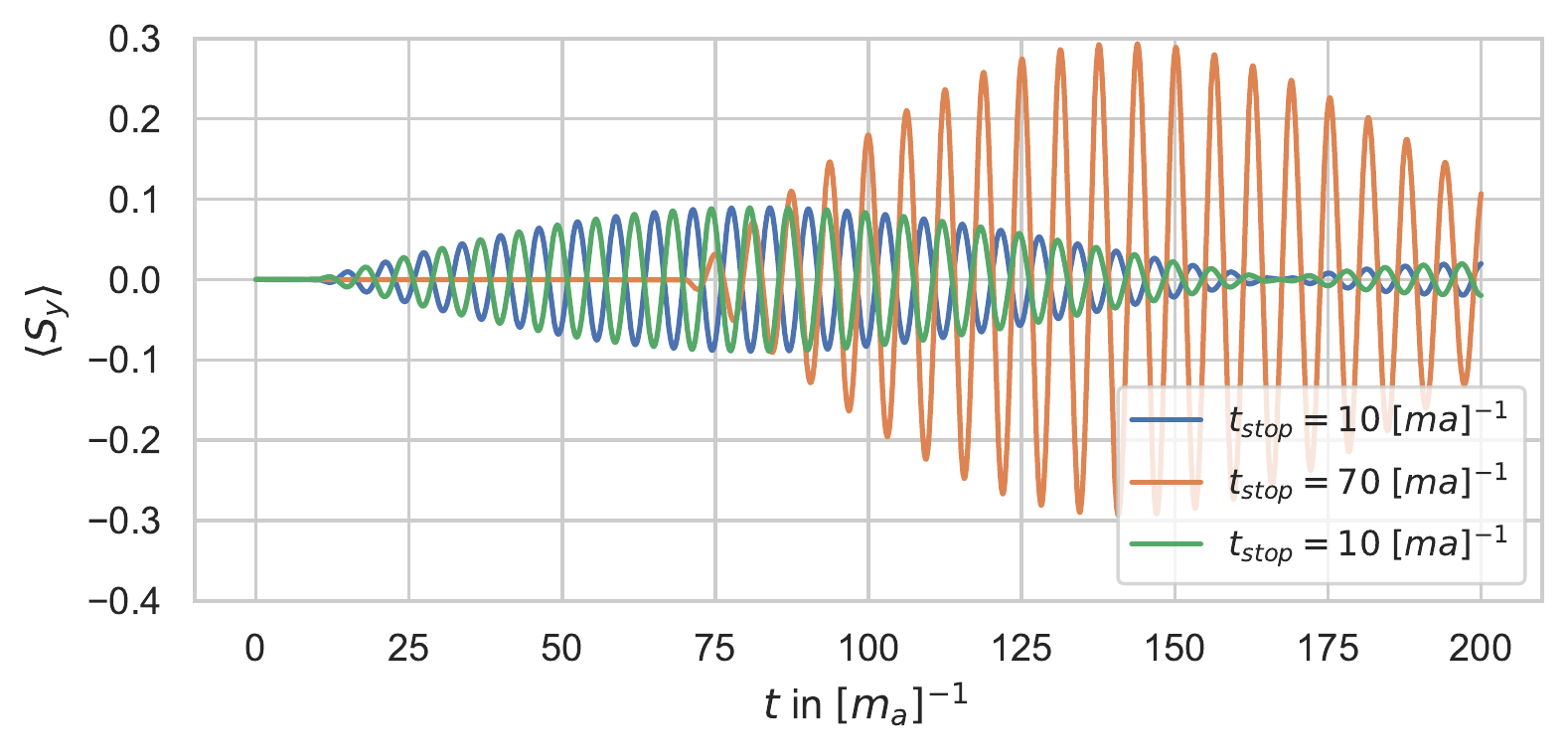}
    \caption{ Evolution of the spin expectation value in the $y$-direction before and after stopping the experiment to allow for spin relaxation or flushing of the spins (the time during the interruption of the measurement is not shown). The blue line corresponds to $t_{stop}= 10\,[m_{a}]^{-1}$ and the orange one to $t_{stop}= 70\,[m_{a}]^{-1}$ with phase $\delta=\delta_{\uparrow}-\delta_{\downarrow}= -1.5$  and the green line corresponds a $t_{stop}= 10\,[m_{a}]^{-1}$ and $\delta= 1.5$.
    }
    \label{fig:spiny_1}
\end{figure}

In Fig. \ref{fig:spiny_1}, we show the transverse spin expectation values for states of the form, Eq.~\eqref{eq:relaxed} after switching on the interaction again. For illustration we do not show the time during which the relaxation takes place but continue directly after $t_{stop}$.

An explicit formula the expectation value of $S_{y}$ is given by,
\begin{equation}
    \langle S_y \rangle = 
    \frac{1}{\sqrt{2}} \sin\left(2 \sqrt{N_a} \omega_{c}t_{stop}\right)
    \sin\left(\sqrt{N_a} \omega_{c} t\right)\cos\left(\sqrt{N_a-1} \omega_{c} t\right)  \cos\left(\delta_{\uparrow}-\delta_{\downarrow}-m_a t\right).
\end{equation}
As we can see, the maximal amplitude of the subsequent oscillations in the spin expectation value is determined by the amplitude of the contribution of the $|N_{a}-1\rangle$ axion state. To get an amplitude $\sim 1$ we therefore need to wait for a time of the order of $\sim 1/(\sqrt{N_{a}}\omega_{c})$ such that there is a significant amplitude for a spin flip in the first running phase.
However, in the more realistic case of a huge number of spins $N_{s}$ we still only need to flip of the order of 1 spin. This reduces the required amount for the initial phase by a factor of $\sim\sqrt{N_{s}}$ as one can directly see from the general Hamiltonian Eq.~\eqref{eq:Hint}.

With the long coherence times, present in experiments like CASPEr, relaxation of the spins may take a long time which would be lost to sensitive measurements. This seems rather inefficient.
Alternatively, one could simply remove the polarized nuclei (or more precisely the corresponding atoms) from the system and fill in a freshly polarized sample, thereby ``renewing'' the spins.

This still seems somewhat wasteful, as it requires flushing the, mostly unused, spins. A more practical way is to change the magnetic field and therefore the Lamor frequency such that a different axion mass is explored. Choosing a suitable time $t_{stop}$ we can then use a single spin sample to ``prime'' the axion state for a whole range of masses. After this priming, the measurement can proceed in the usual way.

\subsection{Measuring the power spectrum and appearance of the axion frequency $\omega_{L}=E\approx m_{a}$}\label{subsec:power}
The essential cause for the vanishing expectation value in an energy eigenstate is that it does not have a physical phase. The ``classical'' field value (see also next section) for the axion $\langle \phi\rangle$ vanishes, not because the typical value of a measurement of it would be zero (or close to it) but because positive and negative values are equally likely.
Slightly more precisely (cf. also Sec.~\ref{sec:classical}),
\begin{equation}
    \left|\phi^{typical}_{measured}\right|\sim \sqrt{\langle \phi^{2}\rangle}\sim\sqrt{N_{a}} \neq 0
\end{equation}
is non-vanishing also for an axion number eigenstate.

\begin{figure}
    \centering
   \includegraphics[width=.8\textwidth]{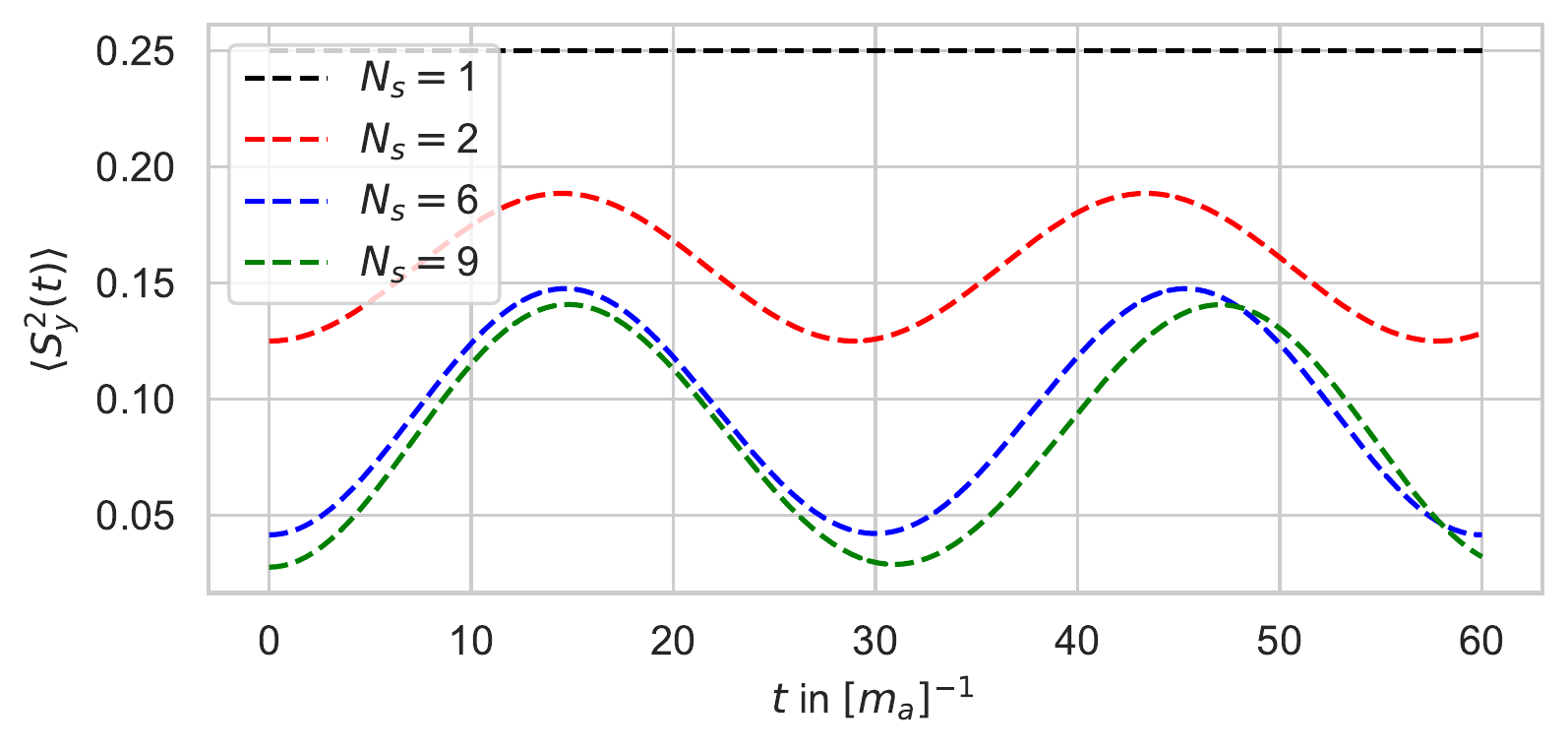}
    \caption{Evolution of the expectation value of the spin squared $\langle (S_{y}(t))^2\rangle$ for an initial state $| N_a, N_s,0 \ket $ with different numbers of spins indicated by the labels and $N_{a}=30$. The coupling frequency used in this figure was $\omega_c = 0.01 \, [m_a]^{-1}$, and the expectation values were normalized with $N_s^{-2}$ for better visibility.}
    \label{fig:squaredspin}
\end{figure}

This suggests measuring a phase insensitive, squared quantity,
\begin{equation}
    \langle S^{2}_{y}(t)\rangle.
\end{equation}
The expectation value for this is plotted in Fig.~\eqref{fig:squaredspin} for a couple of exemplary spin numbers for an initial axion number eigenstate. It is indeed non-vanishing and grows on a time scale $\sim 1/(\sqrt{N_{a}}\omega_{c})$ set by the coupling.
However, it does not feature an oscillation with the Larmor frequency.

Unfortunately, also already at vanishing times the value of $\langle S^{2}_{y}(t)\rangle$ is non-vanishing, i.e. there is at least a part that is clearly unconnected to coupling to the axion. However, this is effectively the problem of measuring the spin values. Indeed, as we can see in Fig.~\ref{fig:squaredspin}, which is normalized to the number of spins $N_s$, this effect relatively decreases with the number of spins as,\footnote{A more precise statement is that $\langle S^{2}_{x}+S^{2}_{y}\rangle/N^{2}_{s}=\langle S^2-S^{2}_{z}\rangle/N^{2}_{s}\geq 1/(2N_{s})$.}
\begin{equation}
    \frac{\langle S^{2}_{y}(t)\rangle}{N^{2}_{s}}\sim \frac{1}{N_{s}}.
\end{equation}
We can therefore benefit from the huge number of spins in the experiment.

\bigskip

\begin{figure}
    \centering
   \includegraphics[width=.8\textwidth]{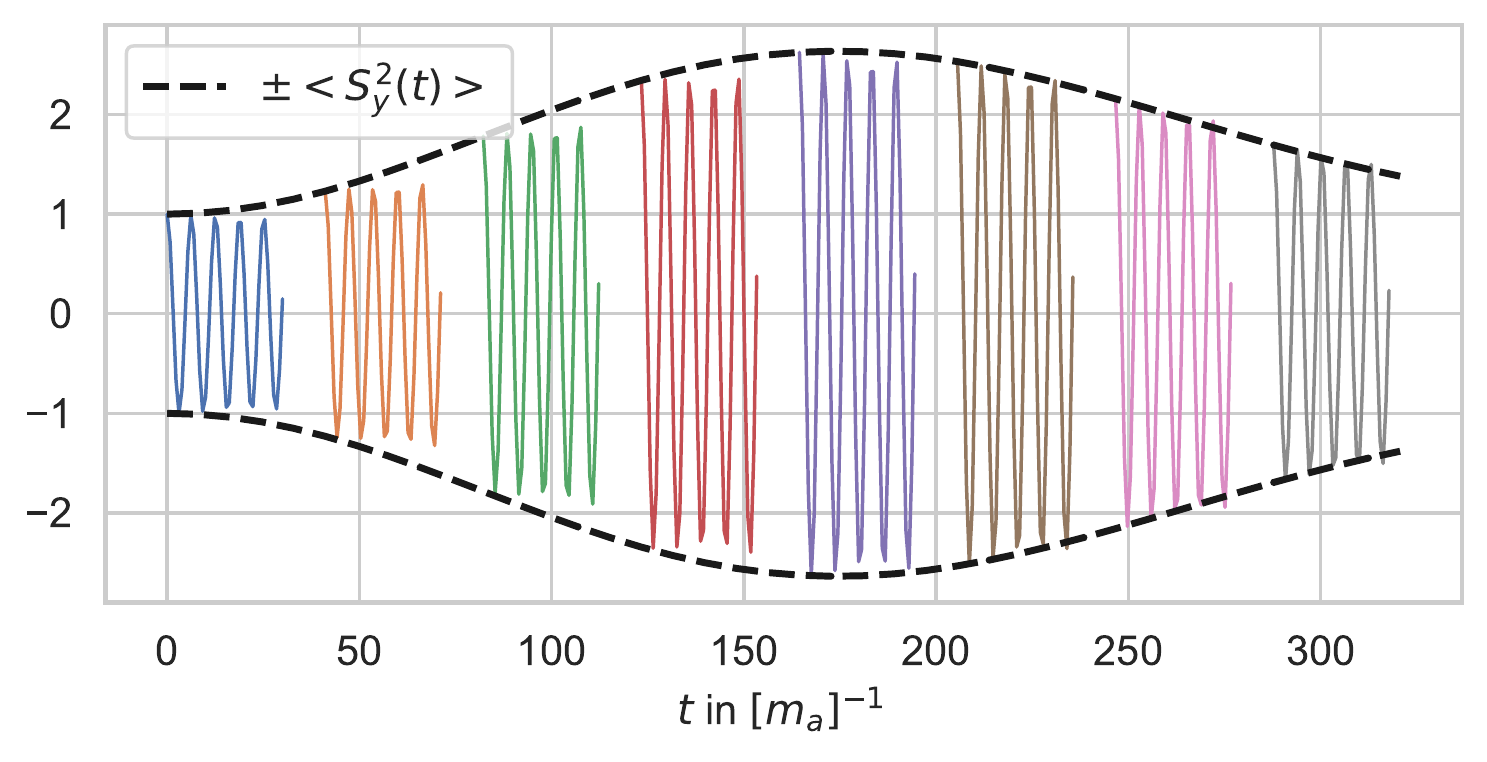}
    \caption{Spin correlator $\langle S_{y}(t'=t+\Delta t)S_{y}(t)\rangle$ for different values of $t'=t$ (starting point of the lines) and a range of $\Delta t=(0-6\pi/\omega_{L})$. The dashed line indicates $\langle (S_{y})^2$ and thereby the envelope of the amplitude of the oscillations. The different colors each denote a different starting point $t = k/8 \cdot 2\pi/(\omega_c \sqrt{\Naxion})$ with $k = 1/40,5/40,\dots$. For better visualization we have chosen a smaller $\omega_c = 0.002 \, [m_a]$ in addition to using $N_a =6$ and $N_s =4$.
    }
    \label{fig:correlator}
\end{figure}

However, this does not yet answer the question about oscillations with frequency $\omega_{L}=E\approx m_{a}$. To find those we can consider the correlator\footnote{For simplicity we consider the symmetrized two point correlator $\frac{1}{2}\langle [S_{y}(t')S_{y}(t)+S_{y}(t)S_{y}(t')]\rangle=\text{Re}\langle S_{y}(t')S_{y}(t)\rangle$ which corresponds to a Hermitean operator.},
\begin{equation}
    \text{Re}\langle S_{y}(t')S_{y}(t)\rangle= \text{Re}\langle S_{y}(t+\Delta t)S_{y}(t)\rangle
\end{equation}
and study its behavior in $\Delta t=(t'-t)$ for different initial times $t$. As shown in Fig.~\ref{fig:correlator} there are indeed oscillations in $\Delta t$ with frequency $\sim \omega_L$.

Using an approximation where we neglect the interaction Hamiltonian $\sim \omega_{c}$ for the evolution between $t$ and $t'$, i.e. for $\sqrt{N_{a}}\omega_{c}\Delta t\ll 1$ we find approximately,
\begin{equation}
    \text{Re}\langle S_{y}(t+\Delta t)S_{y}(t)\rangle\sim \langle S^{2}_{y}(t)\rangle \cos\left(\omega_{L}\Delta t\right).
\end{equation}
This makes it evident that the amplitude of the oscillation in this correlator is directly linked to the expectation value of the square of the transverse spin.

Performing the Fourier transform of the correlator over a small number of these oscillations we can extract an approximation of the spin power spectrum,
\begin{equation}
  P_{S_y}(\omega)[t] \sim  \int^{t+{\rm few}\times 2\pi/\omega_{L}}_{t} d\Delta t \, \text{Re}\langle S_{y}(t+\Delta t)S_{y}(t)\rangle \sim  \langle S^{2}_{y}(t)\rangle \delta_{finite}(\omega-\omega_{L}),
\end{equation}
where $\delta_{finite}$ is an appropriate, approximate, finite time version of the $\delta$-function. Moreover, on the right hand side we have, again, used the approximation with $\sqrt{N_{a}}\omega_{c}\Delta t\ll 1$.
This power spectrum therefore indeed features a peak at $\omega_L$.

The reason that we have to restrict ourselves to a finite range is that due to the continuing interaction with the axions the amplitude of the oscillations in the correlator changes in time. The system is not in a stationary state over long time ranges. 
But, after all, what we want to measure is the axion-induced growth in the amplitude of the oscillations over time. It therefore makes sense to perform the Fourier transform only locally and then consider the growth in the peak at $\omega_L$ of $P_{S_y}(\omega)[t]$ as a function of time. 

While we have indicated taking the Fourier transform only over a small range of a few oscillations, in practice this restriction is rather mild. The evolution in the amplitude occurs on a time scale set by $\sim \sqrt{N_{a}}\omega_{c}$. Taking account of the large hierarchy between $\sqrt{N_{a}}\omega_{c}$ and $\omega_{L}$ there should be ample room to choose 
${\rm few}\ll \omega_{L}/(\sqrt{N_{a}}\omega_{c})$ such that not too much evolution occurs.

For an initial energy eigenstate $|N_{a},N_{s},0\rangle$ and for sufficiently small times $t$ and $\Delta t$ we can compute the correlator in lowest non-trivial order in the coupling frequency,
\begin{equation}
\label{eq:squarespin}
     \text{Re}(\langle S_{y}(t+\Delta t)S_{y}(t)\rangle)= \left[\frac{N_s}{4}  + \frac{\omega_c^2}{2} N_s^2 N_a t^2\right] \cos (m_a \Delta t)\,.
\end{equation}

\section{Comparison to the Classical Treatment}\label{sec:classical}
\subsection{Measuring the expectation value}
In Sec.~\ref{subsec:simpleexample} we have obtained a signal oscillating with the axion frequency $m_{a}$, Eq.~\eqref{eq:simplestateresult}. In particular,
\begin{equation}
\label{eq:spiny}
    \langle S_y \rangle 
    = - \frac{\omega_c \sqrt{\Naxion}\Nspin t}{2} \cos (m_a t).
\end{equation}
Let us now compare this to the result of the classical calculation.

The classical equations of motion are,
\begin{eqnarray}
\label{eq:classsicaleom}
\frac{d\vec{S}}{dt}&=& \omega_{L}\vec{S}\times\frac{\vec{B}}{|\vec{B}|}- 2g_{d}\phi(t,x=0) \vec{S}\times \vec{E}.
\end{eqnarray}

For small values of $g_{d}$ a good strategy is to go to a system that rotates with the Larmor frequency that, in our case, is equal to the axion frequency $m_{a}$, and then solve the equation in this system in linear order in time before rotating back to the original frame. 
This is similar to going to the interaction picture and back in the quantum mechanical situation.

Starting with the initial condition,
\begin{eqnarray}
\phi(t,x=0)&=&\phi_{0}\cos(m_{a}t),\\\nonumber
S_{z}(t=0)&=&S_{0}\\\nonumber
S_{x}&=&S_{y}=0,
\end{eqnarray}
we obtain the following approximate solution,
\begin{equation}
\label{eq:classicalapprox}
S_{y}\approx -g_{d}E_{x}\phi_{0}S_{0}t\cos(m_{a}t).
\end{equation}
Note that, to obtain this we have used that the time average of $\cos^{2}(m_{a}t)$ is equal to $1/2$, which eliminates the factor of $2$ present in Eq.~\eqref{eq:classsicaleom} and corresponds to the rotating wave approximation 
(one could also write $\cos^{2}(m_{a}t)=1/2(1+\cos(2m_{a}t))$ and drop the quickly oscillating term).

This already exhibits a behavior similar to Eq.~\eqref{eq:spiny}.
Let us now also consider the quantitative coefficients.
To do so we can calculate the expectation values and insert them for their classical counterparts.

It is straightforward to check that, for the state given by Eq.~\eqref{eq:superposstate},
\begin{eqnarray}
S_{z}&=&\frac{N_{s}}{2}
\\\nonumber
S_{x}&=&S_{y}=0.
\end{eqnarray}

For the expectation value of the axion field operator with respect to the initial state Eq.~\eqref{eq:superposstate} we find,
\begin{eqnarray}
    \bra \Psi_0 | \phi (t) | \Psi_0 \ket\!\!\! &=&\!\!\! \frac{\varphi(0)}{\sqrt{2m_{a}}}\bra \Psi_0 |a | \Psi_0 \ket e^{im_at} + \frac{\varphi(0)}{\sqrt{2m_{a}}}\bra \Psi_0 |a^{\dagger} | \Psi_0 \ket e^{-im_at}
    = \frac{ \varphi(0) \sqrt{N_a}}{\sqrt{2m_{a}}} \cos(m_at).
\end{eqnarray}

Putting all this together the quantum mechanical Eq.~\eqref{eq:spiny} and the classical calculation Eq.~\eqref{eq:classicalapprox} match, when using the identification Eq.~\eqref{eq:omegac}, $\omega_{c}=g_{d}\varphi(0)/\sqrt{2m_{a}}$.

One can now repeat the exercise for an initial energy eigenstate such as, Eq.~\eqref{eq:energyeigenstate}. The results still match, since the expectation value for $S_{y}$ vanishes but so does the expectation value of the field operator. 
In this sense both states follow the expectation that one would naively\footnote{The right hand side of the equation actually contains $\langle \phi \vec{S}\rangle$. It is an approximation to use instead $\langle \phi\rangle\langle \vec{S} \rangle$.} obtain from the Ehrenfest theorem~\cite{Ehrenfest:1927swx}.

Yet, there is another important ingredient in the standard classical calculation. The amplitude of the field oscillation is obtained from the local mean density $\rho_{a}$ via,
\begin{equation}
\label{eq:classicalfield}
\phi^{cl}_{0}=\sqrt{\frac{2\rho_{a}}{m^2_{a}}}.
\end{equation}

We can now ask whether this is a good expectation in our quantum mechanical state. Indeed, this is where there is a big difference between the two considered states.

The energy density is given by
\begin{equation}
\langle\rho_{a}\rangle =\langle {\mathcal{H}}(x=0)\rangle \approx \langle \frac{1}{2}\left(\dot{\phi}^2+m^{2}_{a}\phi^2\right)\rangle=|\varphi(0)|^2m_{a}\langle a^{\dagger}a\rangle.
\end{equation}
This can be evaluated for the two different states,
\begin{eqnarray}
\label{eq:energyrelationeigenstate}
\langle \rho_{a}\rangle &=&|\varphi(0)|^2m_a \left(N_{a}+\frac{1}{2}\right),\qquad\qquad\qquad\mathrm{Eq.}~\eqref{eq:energyeigenstate}
\\
\langle \rho_{a}\rangle &=&|\varphi(0)|^2m_{a}\left( N_{a}\right),\qquad \qquad\qquad\qquad \,\,\,\,
 \mathrm{Eq.}~\eqref{eq:superposstate}.
\end{eqnarray}

We can now compare the amplitude of the expectation value with the naively obtained classical amplitude,
\begin{eqnarray}
\frac{\langle \phi\rangle}{\phi^{cl}_{0}} &=&0,\qquad\qquad\qquad\mathrm{Eq.}~\eqref{eq:energyeigenstate}
\\
\frac{\langle \phi\rangle}{\phi^{cl}_{0}} &=&\frac{1}{2}\qquad\qquad\qquad \,
 \mathrm{Eq.}~\eqref{eq:superposstate}.
\end{eqnarray}

In the energy eigenstate the expectation value is as different as possible from the naive classical amplitude. While the expectation value for the state Eq.~\eqref{eq:superposstate} is not yet equal to the classical expectation, it is much closer.

Indeed, we could now consider a more complicated state, obtained by superimposing a larger number of states, e.g.
\begin{equation}
    \frac{1}{\sqrt{k+1}}\left(|N_{a}\ket+|N_{a}-1\ket+\ldots |N_{a}-k\ket\right). \label{eq:k-state}
\end{equation}
This has energy density and field expectation values,
\begin{eqnarray}
    \langle\rho_{a}\rangle&=&|\varphi(0)|^2m_{a}\left( N_{a}-\frac{k-1}{2}\right), 
    \\\nonumber 
    \langle \phi\rangle&=&\frac{\varphi(0)}{\sqrt{2m_{a}}}\frac{2}{k+1} \left(\sqrt{N_a}+\sqrt{N_a-1}+\ldots \sqrt{N_a-(k-1)}\right)\approx \frac{\varphi(0)}{\sqrt{2m_{a}}}\frac{2k}{k+1}\sqrt{N_{a}}.
\end{eqnarray}
For large $N_a$ and moderately large $k$ we can then check,
\begin{equation}
\frac{\langle \phi\rangle}{\phi^{cl}_{0}}\approx \frac{k}{k+1} \rightarrow 1. \label{eq:increasing_k}
\end{equation}
Already with the superposition of only a few states we are quite close to the classical expectation.

While the above states are still quite special, we expect that this general tendency holds, i.e. if we superimpose even a modest number of different energy states we obtain a situation that has a result quite close to the classical one. As explained in the previous section such a superposition quite naturally arises when we perform different runs of the experiment at the same frequency and in between change the spin samples.

\bigskip

To get a handle on the expected corrections to the classical result we can start from the quantum mechanical equations of motion and the apply Ehrenfest's theorem with respect to the interaction Hamiltonian to obtain the evolution of the expectation values,
\begin{eqnarray}
    \frac{d}{dt}\left\langle S_{k}\right\rangle&=&i\langle [ H_{int},S_{k}] \rangle = i2\oc\langle (a+a^{\dagger}) [S_x,S_k] \rangle  
    = -2\oc\varepsilon_{xkl}\langle (a+a^{\dagger}) S_l \rangle 
   \\\nonumber
    &\approx& -2\oc\varepsilon_{xkl}\langle a + a^{\dagger} \rangle \langle S_l \rangle.
\end{eqnarray}

In the second line we have factorized the expectation value of a product into a product of expectation values. In general this is only an approximation. The validity of this approximation can be quantified by
\begin{equation}
    \frac{\langle \phi S_k\rangle - \langle \phi \rangle \langle S_k \rangle}{\langle \phi S_k \rangle}. \label{eq:correlator}
\end{equation}
In the case of only one spin, we can calculate this quantity analytically for the example state $\frac{1}{\sqrt{2}}(|\Naxion\ket + |\Naxion-1\ket)|1,0\ket$. This is possible because the time evolution does not mix the subspaces of the states making up the superposition, the measurements do. We then have for the above correlator in the limit $\Naxion \gg 1$:
\begin{eqnarray} 
    \bigg| \frac{\langle \phi S_z\rangle - \langle \phi \rangle \langle S_z \rangle}{\langle \phi S_z \rangle} \bigg|&\approx& 0\nonumber\\
    \bigg|\frac{\langle \phi S_y\rangle - \langle \phi \rangle \langle S_y \rangle}{\langle \phi S_y \rangle} \bigg|&\approx& \frac{1}{4}(3-\cos(2m_at)) \leq 1.
\end{eqnarray}
Furthermore, $\langle (a+a^{\dagger})S_x \rangle =0$, which is consistent with the classical equations of motion.

\begin{figure}
\centering
\includegraphics[width=.8\textwidth]{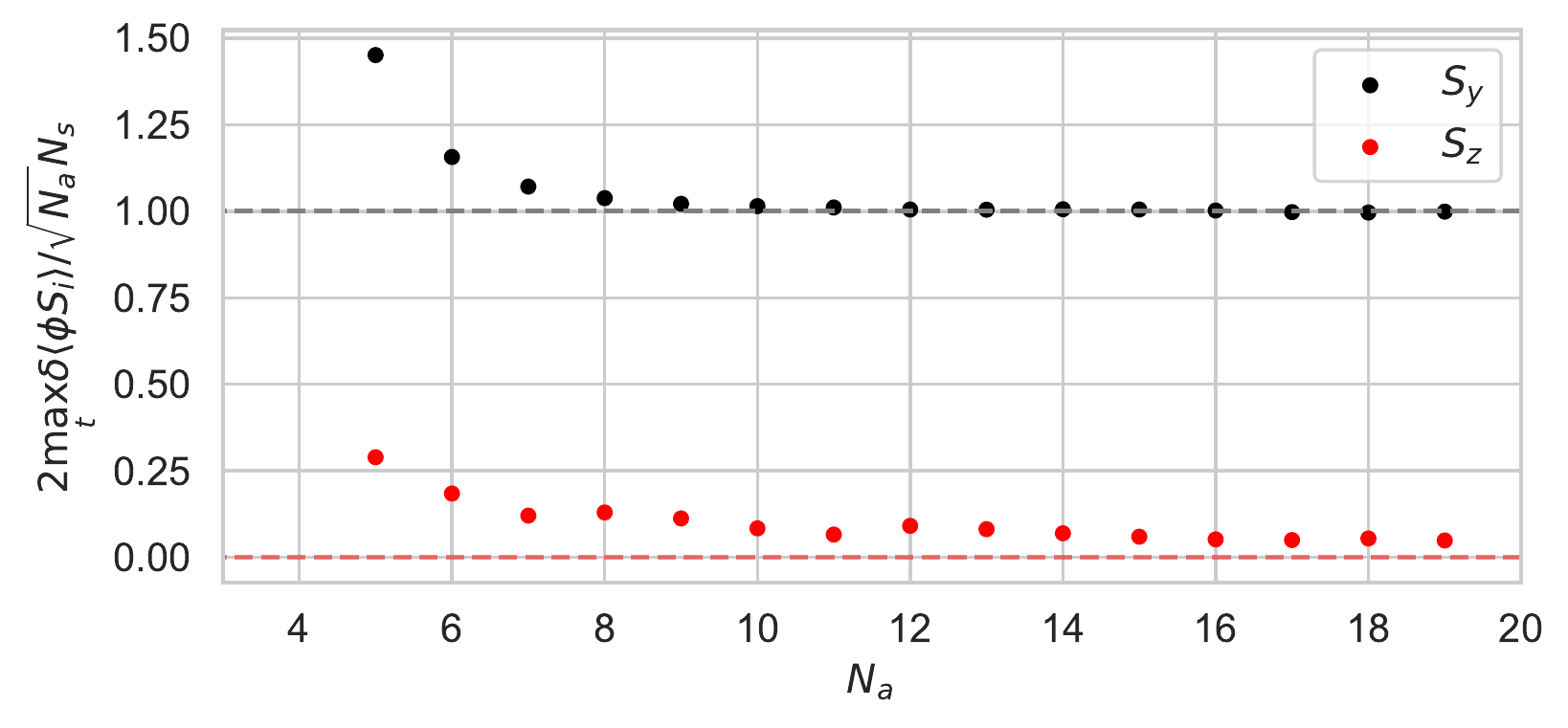}
\caption{Plot of the maximal values of $\delta \langle \phi S_i \rangle = \big|\langle \phi S_i \rangle - \langle \phi \rangle \langle S_i \rangle \big|$ normalized to $\max_t \langle \phi S_i \rangle = \sqrt{N_a}\frac{N_s}{2}$ for $i = y,z$ and $5 \leq N_a \leq 100$. This normalisation indeed gives an upper bound on the correlators, as $\delta \langle \phi S_i \rangle$ takes on its maximum value at the same time $t$ as  $\langle \phi S_i \rangle$ The number of spins for this plot was chosen as $N_s = 4$, and $\omega_c = 0.01 \, [m_a]$. We chose the maximal time for the time evolution to be $T = 2 \, [\omega_c]^{-1}$. The results of this were obtained numerically using Qutip \cite{JOHANSSON20121760,JOHANSSON20131234}.}
\label{fig:correlators}
 \end{figure}

This suggests that the evolution of $S_{y}$ (in the rotating frame)  is reasonable well approximated by the classical evolution that is mostly driven by $S_{z}$, whereas the approximation is less good for the evolution of $S_{z}$. Of course after a sufficiently long evolution the latter then also worsens the approximation for $S_y$.
To show that this is not an artifact of the choice $N_s=1$, we carried out a numerical calculation of these correlators for $N_s = 4$ shown in \ref{fig:correlators}.  

\begin{figure}
 \centering
 \includegraphics[width=.8\textwidth]{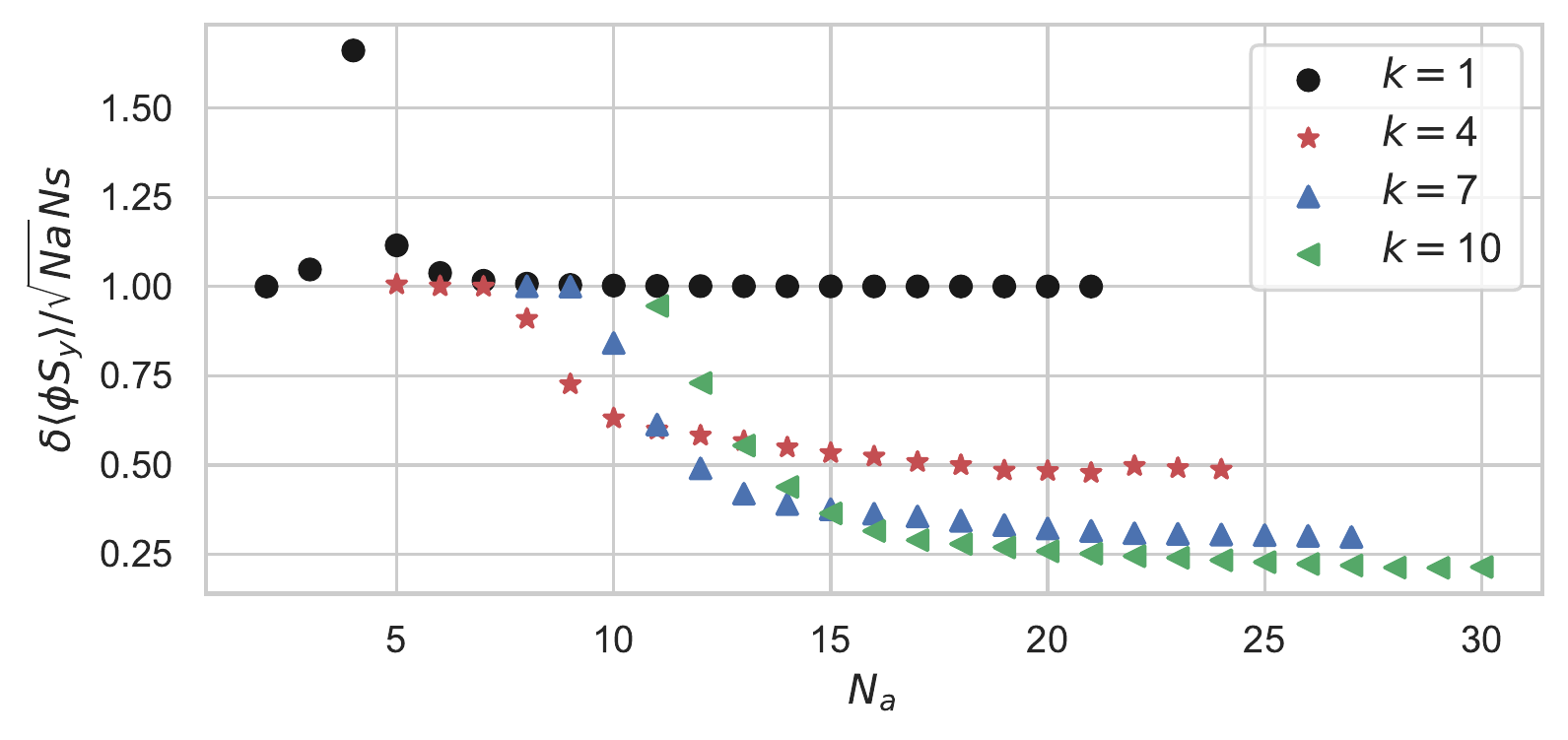}
 \caption{Plot of the maximal values of $\delta \langle \phi S_i \rangle = \big|\langle \phi S_y \rangle - \langle \phi \rangle \langle S_i \rangle \big|$ normalized to $\max_t \langle \phi S_y \rangle = \sqrt{N_a}\frac{N_s}{2}$ for $k+1 \leq N_a \leq k+20$, where different values of $k$ denote different initial states of the type \ref{eq:k-state}. The number of spins for this plot was chosen as $N_s = 4$, and $\omega_c = 0.01 \, [m_a]$. We chose the maximal time for the time evolution to be $T = 2 \, [\omega_c]^{-1}$. The results of this were obtained numerically using Qutip \cite{JOHANSSON20121760,JOHANSSON20131234}.\label{fig:increasingly_coherent}}
 \end{figure}

 We expect the size of the deviation induced by the factorisation to be dependent on the initial state. As mentioned before, 
 using the initial state \eqref{eq:k-state} for $k=1$ leads to significant differences between the classical and quantum picture \eqref{eq:superposstate}, which on the other hand decrease with increasing $k$ \eqref{eq:increasing_k}. In Fig.~\ref{fig:increasingly_coherent} it can be seen that this is also the case for the correlator \ref{eq:correlator} with respect to $S_y$ (as long as we choose $k\lesssim \sqrt{N_{a}}$ which roughly gives the range for which states of the form \eqref{eq:k-state} come closer to the coherent state).

 \begin{figure}
\centering
\includegraphics[width=\textwidth]{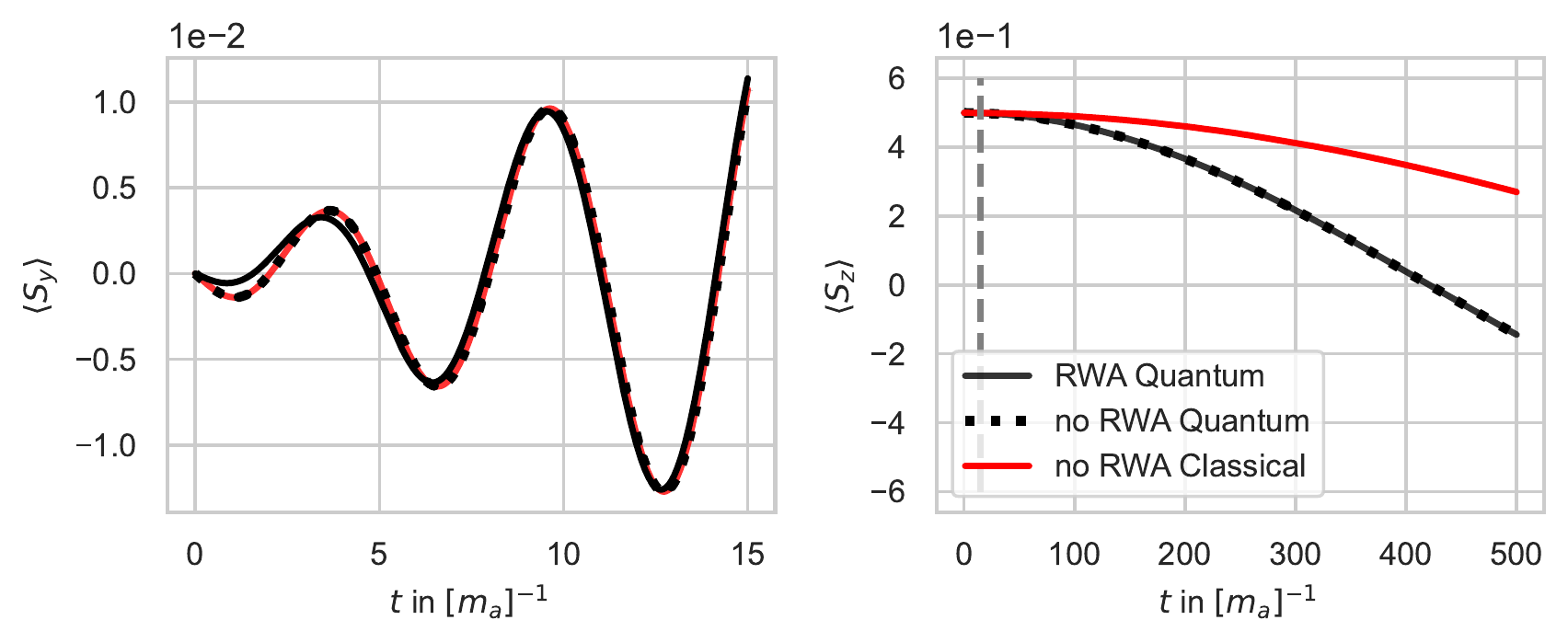}
\caption{Plot showing the difference of different results at different time scales. We see that the difference due to the RWA can be seen easiest at early times, whereas the difference between quantum and classical calculations becomes visible at later times. The values chosen for this are $N_a = 4$, $N_s =1$ and $\omega_c = 0.001 \, [m_a]^{-1}$. Here the grey dashed line in the second plot denotes the end of the time frame of the first plot. The results of this were obtained numerically using Qutip \cite{JOHANSSON20121760,JOHANSSON20131234}.
}
\label{fig:timescales}
 \end{figure}

As already suggested, corrections due to the factorisation of expectation values in the classical eom affect the long-time behaviour of classical solutions, occurring on time-scales set by the interactions $\sim \omega_c$. Meanwhile, the RWA affects the early time behaviour of the spin evolution. In  \ref{fig:timescales} we can see that the classical solution without the RWA yields a better fit to the quantum mechanical result at early times when compared with the same quantum mechanical simulation with the RWA applied. The deviations induced by the rotating wave approximations become negligible after some oscillations of the axion field, whereas the classical result starts showing significant deviations from the quantum mechanical result at timescales $\sim \omega_c^{-1}$.

\bigskip

An additional approximation in the classical dynamics is due to the ansatz we have used for the axion field. So far, we have neglected the effects of axion number non-conservation when deriving the classical solution~\eqref{eq:classicalapprox}, meaning we neglect the backreaction of the spin onto the axion field amplitude. This can be included by also fully accounting for the evolution of the axion field.
As typically $N_{s}\ll N_{a}$ and indeed during the runtime of the experiment only a small fraction of the spins are flipped, in most cases this effect is rather small. We nevertheless briefly discuss it in Appendix~\ref{app:backreaction}.

\subsection{Measuring the power spectrum}\label{subsec:powerclassical}
In Sec.~\ref{subsec:power} we have proposed to measure the two point correlator $\text{Re}\langle S_{y}(t+\Delta t)S_{y}(t)\rangle$ and extract from this the power spectrum, instead of measuring the simple expectation value.

Based on the results obtained in Sec.~\ref{subsec:power} we can now make a similar comparison between the classical and the quantum mechanical result for the expectation value of the energy density in the axion field and the amplitude of the correlator as done in the previous subsection.

Using Eq.~\eqref{eq:squarespin} for the spin correlator in lowest non-trivial order in $\omega_{c}t$, the relation Eq.~\eqref{eq:energyrelationeigenstate} for the energy density in an energy eigenstate, as well as the coupling relation Eq.~\eqref{eq:classicalfield} we find that for an initial energy eigenstate $|N_{a},N_{s},0\rangle$,
\begin{equation}
\label{eq:quantumenergydep}
    \langle S^{2}_{y}(t)\rangle=  \frac{1}{4}N_{s} + N_{s}^2\frac{g^{2}_{d}E^{2}_{x}}{4 m_a}\left(\frac{\rho_a}{m_a} - \frac{|\varphi(0)|^2}{2}\right) t^2.
\end{equation}

This can now be compared to the classical value that can be obtained by simply squaring Eq.~\eqref{eq:classicalapprox} and using Eq.~\eqref{eq:classicalfield} to relate to the density. Finally, inserting Eq.~\eqref{eq:omegac} we have,
\begin{equation}
\label{eq:classicalspin}
    S^{2}_{y,cl}=N^{2}_{s} \frac{g^{2}_{d}E^{2}_{x}}{2m_a}\frac{\rho_{a}}{m_{a}} t^2.
\end{equation}
Comparing Eqs.~\eqref{eq:quantumenergydep} and~\eqref{eq:classicalspin} we find that for large $N_{a}$ and $N_{s}$ they agree up to a factor of $2$. The latter, however, reduces to $1$ if we average over the, a priori unknown, initial phase, $\psi$, of the classical field,
\begin{equation}
    S_{y,cl}(t)=S_{y,cl}\cos(\omega_{L}t+\psi).
\end{equation}
Denoting the average over the phase $\psi$ by $\langle\cdot\rangle_{\psi}$ we then have,
\begin{eqnarray}
\nonumber
 &&\!\!\!\!\!\!\!\!\!\!\!\!\!\!\!\!\!\!   \langle S_{y,cl}(t+\Delta t)S_{y,cl}(t)\rangle_{\psi}
   \\ \nonumber
    &&\qquad=S^{2}_{y,cl}\langle \cos^{2}(\omega_{L}t+\psi)\cos(\omega_{L}\Delta t)-\cos(\omega_{L}t+\psi)\sin(\omega_{L}t+\psi)\sin(\omega_{L}\Delta t)\rangle_{\psi}
    \\
    &&\qquad=\frac{1}{2}S^{2}_{y,cl}\cos(\omega_{L}\Delta t).
\end{eqnarray}

In addition we have small corrections from the always non-vanishing spin fluctuations $\sim N_{s}/4$ and from the ``vacuum contribution'' to the energy density.

\section{Brief summary}\label{sec:conclusions}

In this note we have clarified the origin of the relevant oscillation frequencies in axion dark matter experiments from a quantum mechanical perspective.
As a simple but pertinent example we have considered a coupling of axions to spins as probed by the CASPEr experiment~\cite{Budker:2013hfa}. We do so by showing that the original quantum field theoretical model and the experimental situation can be well approximated by a Jaynes-Cummings type model~\cite{Jaynes:1963zz} and then discussing suitable observables.\footnote{Once more we stress, that our aim is to simply discuss the specific case of axion experiments and not to add anything new to the general discussion of the Jaynes-Cummings model or the quantum/classical connection.}

In our concrete case of interest, the CASPEr experiment, we have three base frequencies: the axion energy $E\approx m_{a}$, the Larmor frequency $\omega_{L}=g\mu_{N}B$ and the coupling frequency due to the elecric dipole moment $\omega_{c}=g_{d}E_{x}\varphi(0)/\sqrt{2m_{a}}$.
For a resonant setup the experimental parameters (in particular the magnetic field) are chosen such that $\omega_L=E\approx m_{a}$. 

But how does this lead to the signal that is a magnetization oscillating with the Larmor=axion frequency?
In the classical field theory calculation it is simply the axion field oscillating with this frequency (and in resonance with the spin precession) that sets the oscillation of the magnetization with $m_{a}$.

From a quantum mechanical perspective we can consider energy/axion number eigenstates with a given spin value in the magnetic field direction.
For example $|N_{a},N_{s}\uparrow, 0\downarrow\rangle$. The interaction Hamiltonian then can flip one spin at the cost of one axion, e.g. to \mbox{$|N_{a}-1,(N_{s}-1)\uparrow, 1\downarrow\rangle$}. This implements energy conservation in the sense that the energy required to flip one spin $\omega_L$ is provided by the absorption of one axion\footnote{Axion number is not conserved by the interaction.}\footnote{In principle the interaction is completely symmetric with regards to absorption/emission of axions. The preference for absorption is due to the fact that we have chose an initial situation where all spins are in the energetically lower state and are therefore ``ready'' to absorb an axion.} with energy $m_{a}=\omega_L$.\footnote{If the resonance condition is not fulfilled, transitions are still possible, but suppressed. Energy conservation is, of course, still valid and can be understood from the fact that $|N_{a},N_{s}\uparrow, 0\downarrow\rangle$ is not an energy eigenstate of the Hamiltonian including the dipole interaction.} 

One can easily check (and we have done so explicitly above) that starting from an energy eigenstate we get an oscillation of the expectation value of the spin in the direction of the magnetic field determined by the small coupling frequency $\sim \omega_c$.
This slow oscillation can be understood from the fact that the dipole interaction splits the previously degenerate set of eigenstates $|N_{a},(N_{s}-N_{d})\uparrow,N_{d}\downarrow\rangle$, $N_{d}=0,\ldots,N_{s}$, by amounts $\sim \omega_c$. Once the interaction is included (e.g. the electric field turned on) an initial state such as $|N_{a},N_{s}\uparrow, 0\downarrow\rangle$ is thus a superposition of states with different energies of the full Hamiltonian and the beating of the different frequencies causes the time evolution with frequency $\sim \omega_{c}$.

However, for an initial energy/axion number eigenstate we do not obtain an oscillating expectation value of the spin in the direction transverse to the magnetic field.
This might be worrying\footnote{Although we note that we currently do not see any strong processes that would have the tendency to create an axion number eigenstate in the first place.} since this is the sought after observable for axion dark matter detection. 
However, this is rectified, once we start from initial states that are not energy eigenstates of the Hamiltonian (without interactions). As we have shown by example, already simple superpositions lead to an oscillation of the spin in the transverse direction of a magnitude similar to that expected from the classical calculation.
The oscillation frequency $\omega_L\approx m_{a}$ arises because this corresponds to the level spacing between the different energy eigenstates of the Hamiltonian without the interaction, which is classified by the axion number (level spacing set by $m_{a}$) and the spin in the magnetic field direction (level spacing by $\omega_L$). 

Even starting from an energy/axion number eigenstate, the experiment itself typically changes the system away from the initial axion number eigenstate. Running the system for a suitable time and re-initializing the spin state then puts the system into a state that exhibits oscillations close to the classical result. A simpler alternative is to measure a suitable correlator of the transverse spin such as $\text{Re}\langle S_y(t+\Delta t)S_y(t)\rangle$ which features oscillations with frequency $\sim \omega_L$ in $\Delta t$ even for an initial axion number state. In other words, the axion effect can be found in the spin power spectrum at a frequency $\sim \omega_L\approx m_a$.

TLDR, for many initial states the quantum mechanical calculation of the observable signal agrees well with the one from classical field theory. While special states that do not well approximate the classcial result exist, they can be modified by a suitable experimental procedure such that a signal which is relatively close to the classical one re-appears. Measuring, the power spectrum works even for special states.

\section*{Acknowledgements}
Once more we would like to thank Dima Budker for starting this project by asking the questions addressed in this note as well as for valuable comments on the manuscript. Moreover, we are grateful to Yevgeny Stadnik and Arne Wickenbrock for useful comments and discussions. 
We are happy to acknowledge that this project has received support from the European Union’s Horizon 2020 research and innovation programme under the Marie Sklodowska-Curie grant agreement No 860881-HIDDeN.

\newpage
\appendix
\section{Comments on effects of measurements on the state}\label{app:measurement}
In this Appendix we want to collect a couple of examples and remarks on the influence of the measurement process on the system and its time evolution after the measurement.

Let us start with a rather general statement.
Measuring the spin in the $x$- or $y$-direction, as done in an experiment like CASPEr, affects the state of the system. This is evident from the non-vanishing commutator 
\begin{equation}
    [H, S_y] = im_{a}S_{x} + 2\omega_{c} S_{z}(a^\dagger + a),\
\end{equation}
that ensures that simultaneous eigenstates do not exist.
This holds even in the absence of interactions $\sim \omega_{c}$.

In particular, if we start from an energy eigenstate, after the measurement the system is usually not an energy eigenstate anymore, but contains contributions from states whose energies differ by amounts $\sim m_{a}=\omega_L$.\footnote{This also holds in the absence of interactions, i.e. for $\omega_c=0$.}
This already resembles the situation needed to obtain observable spin expectation values that oscillate with frequency $\omega_L$.
However, depending on the measurement procedure this does not automatically ensure that the expectation value of the field behaves in a ``close to classical'' manner (measuring the power spectrum as suggested in Sec.~\ref{subsec:power} seems much safer in regard to this).

In the following we will give two explicit examples of the side effects of different measurement procedures. Looking what happens in a more realistic modelling of the measurement procedure as implemented in CASPEr would be interesting but is beyond the scope of the present note.

\subsubsection*{Complete spin measurement}
The most basic way to implement a measurement in quantum mechanics is to do a full measurement of a Hermitean operator. This returns an eigenvalue of the operator and after the measurement the state is in the corresponding eigenstate.

Let us now do this for the simplest case of only one spin. Using the initial state $|\Naxion,1,0\rangle$ we can simply use the result for the time evolved state Eq.~\eqref{eq:res-Ns=1} explicitly writing the state vectors,
\begin{equation}
    |\Psi(t) \ket = \cos\left(\sqrt{N_{a}}\omega_{c}t\right) |\Naxion \rangle \otimes |\uparrow \rangle +i \sin\left(\sqrt{N_{a}}\omega_{c}t\right)|\Naxion-1\rangle \otimes |\downarrow \rangle \,.
\end{equation}
We can rewrite this into the spin states in the $y$-direction (or alternatively $x$-direction) using,
\begin{align}
    |\uparrow \ket &= \frac{1}{\sqrt{2}}(|\rightarrow \ket + | \leftarrow \ket) \nonumber\\
    | \downarrow \ket &= -\frac{i}{\sqrt{2}}(|\rightarrow \ket - | \leftarrow \ket) \,
\end{align}
such that
\begin{align}
    |\Psi(t) \ket &= \frac{1}{\sqrt{2}}(\cos\left(\sqrt{N_{a}}\omega_{c}t\right)|\Naxion \rangle + \sin\left(\sqrt{N_{a}}\omega_{c}t\right)|\Naxion-1\rangle)\otimes|\rightarrow \ket   \\\nonumber
    &+ \frac{1}{\sqrt{2}}(\cos\left(\sqrt{N_{a}}\omega_{c}t\right)|\Naxion \rangle - \sin\left(\sqrt{N_{a}}\omega_{c}t\right)|\Naxion-1\rangle)\otimes|\leftarrow \ket.
\end{align}
Now, at time $t_{meas}$ perform a measurement in the $y$-direction of the spin. At this point the system is projected onto an eigenstate of the spin in the $y$-direction. Let us consider the example where the spin points in the right direction. To describe the state after the measurement we have to project the state onto this direction using the projection operator,
\begin{equation}
    P^{\rightarrow}_{y}=|\rightarrow\rangle\langle\rightarrow|.
\end{equation}
This yields
\begin{equation}
    P^{+}_y |\Psi(t_{meas}) \ket =\frac{1}{\sqrt{2}}(\cos\left(\sqrt{N_{a}}\omega_{c}t_{meas}\right)|\Naxion, \rightarrow \ket  + \frac{1}{\sqrt{2}}(\sin\left(\sqrt{N_{a}}\omega_{c}t_{meas}\right)|\Naxion-1, \rightarrow \ket. 
\end{equation}
Simple projection gives a not yet normalized state. However, this can be easily remedied because, at least immediately after the measurement of a spin in the right direction, we have a 100\% probability to be in this state. Therefore, we can simply normalize to $1$,
\begin{eqnarray}
   |\Psi_{\rm new}\ket&=&\frac{1}{{\rm \sqrt{1/2}}} P^{+}_y |\Psi(t_{meas}) \ket 
   \\\nonumber
   &=&(\cos\left(\sqrt{N_{a}}\omega_{c}t_{meas}\right)|\Naxion, \rightarrow \ket  + (\sin\left(\sqrt{N_{a}}\omega_{c}t_{meas}\right)|\Naxion-1, \rightarrow \ket. 
\end{eqnarray}
To answer the question what we will measure in subsequent measurements at later times we can now perform a further time evolution of this state $|\Psi_{\rm new}\ket$. To do so we go back into the initial basis,
\begin{align}
\label{eq:meas}
  |\Psi_{\rm new}\ket& = \frac{1}{\sqrt{2}}(\cos\left(\sqrt{N_{a}}\omega_{c}t_{meas}\right)(|\Naxion, \uparrow \ket + i |\Naxion, \downarrow \ket ) \\\nonumber
  &- \frac{i}{\sqrt{2}}(\sin\left(\sqrt{N_{a}}\omega_{c}t_{meas}\right)(|\Naxion-1, \uparrow \ket - i |\Naxion-1, \downarrow \ket).
\end{align}

\begin{figure}
    \centering
   \includegraphics[width=.8\textwidth]{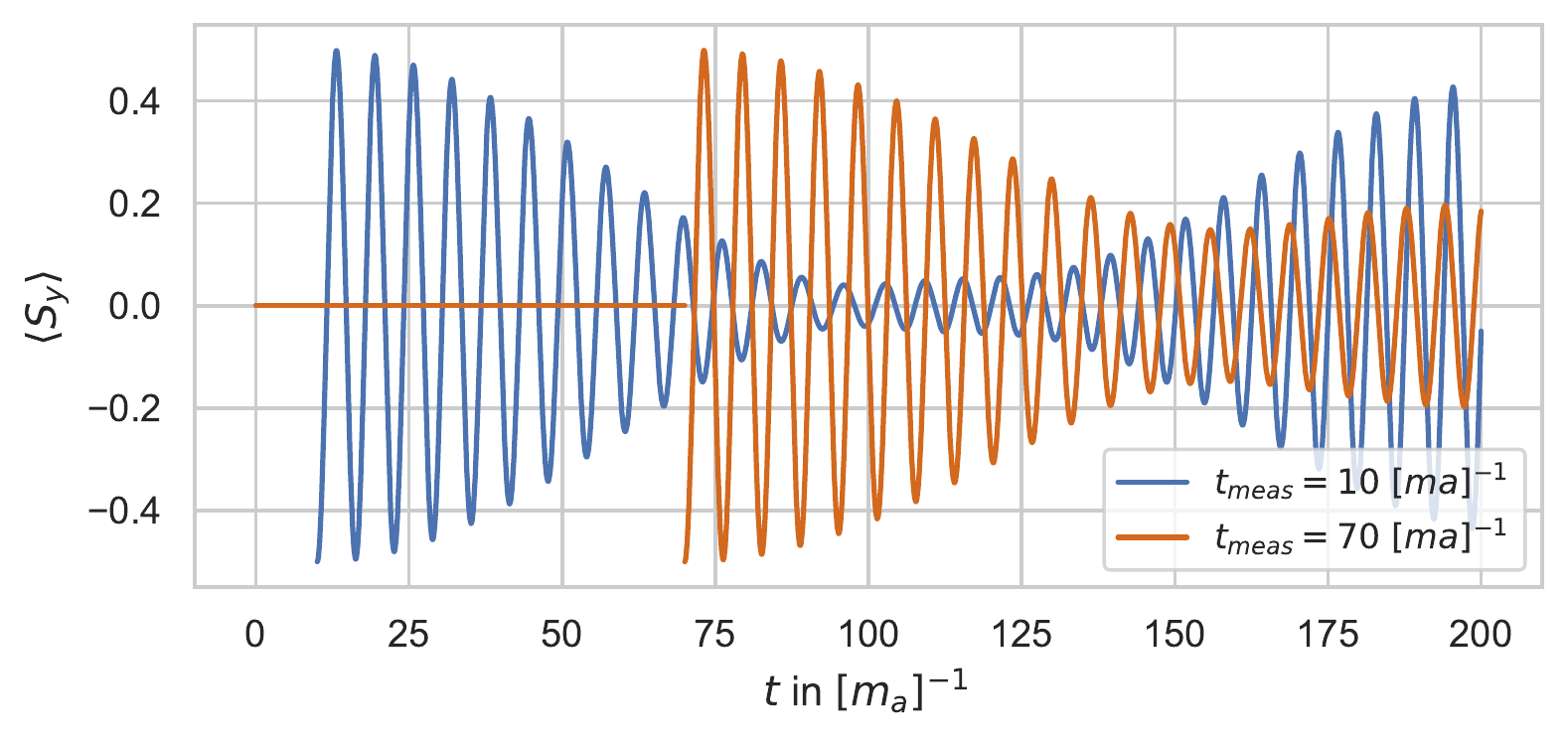}
    \caption{Evolution of the expectation value in y-direction of the spin before and after the measurement, for $\Nspin = 1$ and $\Naxion =2$. The blue line corresponds to $t_{meas}= 10\,[m_{a}]^{-1}$ and the orange one to $t_{meas}= 70\,[m_{a}]^{-1}$. }
    \label{fig:spinmeas}
\end{figure}

At this point we can already clearly see that this is not an energy eigenstate of the Hamiltonian without the dipole interaction. We therefore expect it to have non-trivial evolution with the axion/Larmor frequency as observed in Sec.~\ref{subsec:simpleexample}. The oscillations can be seen in Fig.~\ref{fig:spinmeas} where we plot the time evolution of the spin expectation value after the measurement.
This is the good news. 

However, as we can also see from Fig.~\ref{fig:spinmeas}, the size of the initial oscillation is independent of the time of the measurement. This is because the measured spin value in the $y$-direction is now also the expectation value $\langle S_{y}\rangle$ which (approximately) evolves according to the classical evolution equation. The axion effect is, therefore, not immediately visible in the initial amplitude of oscillation of the expectation value.

One may wonder whether this is an effect of the single spin system and of measuring a non-vanishing spin in the $y$-direction.
However, even for more than one spin this type of measurement may have undesirable effects. For example, for an even number of spins we can indeed measure zero spin in the $y$-direction. However, if we consider the subsequent time evolution of the expectation vale $\langle S_{y}\rangle$ the axion effect is also not present in the classical way. The reason is that in this case the expectation value, $\langle S_{z}\rangle$, of the spin in the $z$-direction vanishes (to make this argument we found the general discussion of higher spin representations in~\cite{Wheeler} very useful). And, classically, it is the latter that drives the leading order evolution of the spin into the transverse plane.

\subsubsection*{Weak/Nondemolition Measurement of spin}

As we have seen, in the standard approach to  measurement, completely measuring the spin in one particular direction, we may rather dramatically affect the state during the measurement process, e.g. destroying the initial $\langle S_z\rangle$ that naively drives the classical evolution. The evolution, after such a measurement is sometimes affected more by the result of the measurement than the axion induced evolution before the measurement.

An alternative is to perform a so-called non-demolition  (cf., e.g.,~\cite{gerry_knight_2004}) 
or weak measurement (cf., e.g.,~\cite{PhysRevD.40.2112}) of the spin,
\begin{equation}
\label{eq:wm}
    |\Psi_{\rm after\,\, measurement}\ket = e^{-i\epsilon S_y} |\Psi\ket.
\end{equation}

\begin{figure}
    \centering
   \includegraphics[width=.8\textwidth]{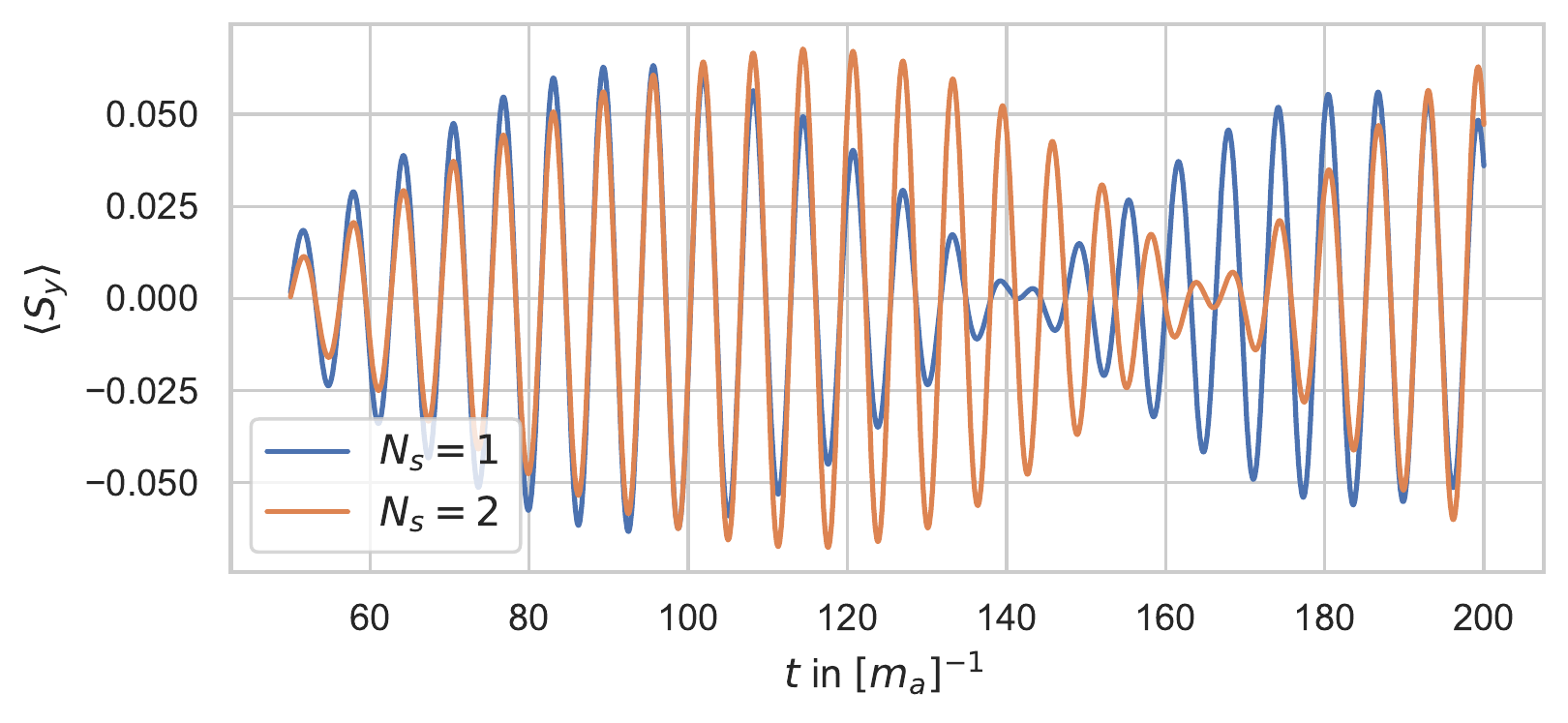}
    \caption{Time evolution of the expectation value of $\langle S_{y}\rangle$ after  weak measurement~\eqref{eq:wm} at time $t_{meas}=50 \,[m_{a}]^{-1}$ for one (blue line) and two spins (orange line) }
    \label{fig:spinmeas2}
\end{figure}

In Fig.~\ref{fig:spinmeas2} we plot the time evolution of the expectation value of $\langle S_{y}\rangle$ after such a weak measurement for an initial energy eigenstate.
As we can see there is a similar problem as in the case of the complete spin measurement in the single spin case. The measurement induces an oscillation, albeit a smaller one, that is not directly linked to the axion effect as it also appears when doing a measurement directly after starting the experiment, i.e. when the axion effect should be negligible.
Even worse, the maximal oscillation amplitude is typically also linked to the strength of the measurement.

In part this may be an artefact performing only a single weak measurement as well as only having a small number of spins. Nevertheless the two example measurement procedures show that, when considering only the simple expectation value of the field, it is non-trivial to obtain a close to classical evolution for an energy eigenstate by just performing spin measurements on the system.
In this sense measuring the power spectrum as discussed in Sec.~\ref{subsec:power} is preferable.

\section{The Classical Equations of Motion with Backreaction}\label{app:backreaction}
To complete our comparison between the quantum and the classical calculation we now also include this backreaction in the classical case\footnote{We retain our simplification of including only one axion energy state.},
In practice, this can be done by including the spin dependent source term on the right hand side of the axion equations of motion
\begin{equation}
    (\square+m^{2}_{a}) \phi = -2 g_d E_x S_x(t) \delta(x).
\end{equation}

We can turn this into a purely time-dependent differential equation by using the mode expansion Eq.~\eqref{eq:modeexpansion} into energy eigenfunctions (of the Klein-Gordon equation). Neglecting all modes but $E_n \approx m_a$ results in the mode equation
\begin{align}
    (\ddot{a} - im_a \dot{a})e^{-imat} + (\ddot{a}^* +im_a \dot{a}^*)e^{im_at} = -4m_a\omega_c S_x(t). \label{eq:modeequation}
\end{align}

We can now define a field $\eta (t) = a e^{-im_at} + h.c.$, such that the equation above becomes
\begin{equation}
    \ddot{\eta} + m_a^2 \eta = -4m_a\omega_c S_x(t).
\end{equation}

This can be coupled to the classical equations of motion of the spin evolution, giving a $5$-dimensional nonlinear coupled system of ODEs. Explicitly, this is given by
\begin{equation}
\frac{\text{d}}{\text{d}t}
\begin{pmatrix}
S_x
\\
S_y
\\
S_z
\\
\dot{\eta}\\
\eta
\end{pmatrix} 
= 
\begin{pmatrix}
\omega_L S_y
\\ 
-\omega_LS_x -2 \oc\eta S_z 
\\
2\oc \eta S_y 
\\ 
-4m_a\oc S_x - m_a^2 \eta
\\ 
\dot{\eta}
    \end{pmatrix}
\end{equation}

When setting the coupling frequency to zero in the equations for $\eta$, the axion field decouples from the spin, and we obtain the approximation in which there is no backreaction on the axion field. 

This approximation remains valid as long as we can neglect the spin term, i.e.,
\begin{equation}    
    \ddot{\eta} = -\left( 4 \frac{\omega_c}{m_a} S_x + \eta \right) m_a^2  \approx -m_a^2 \eta.
\end{equation}

Note that, at small times, this is already satisfied for moderately large axion numbers, as the contribution of the source term is suppressed by the hierarchy $\omega_c/m_a \ll 1$. Therefore, effects due to the backreaction typically only appear on longer time scales.

\bibliographystyle{utphys}
\bibliography{references}

\end{document}